\documentclass[longbibliography,prb,twocolumn,superscriptaddress,amssymb]{revtex4-2}
\usepackage[english]{babel}
\usepackage[colorinlistoftodos, color=green!40, prependcaption]{todonotes}
\usepackage{amsthm}
\usepackage{mathtools}
\usepackage{physics}
\usepackage{xcolor}
\usepackage{graphicx}
\usepackage{adjustbox}
\usepackage{placeins}
\usepackage[T1]{fontenc}
\usepackage{lipsum}
\usepackage{csquotes}
\usepackage[pdftex, pdftitle={Article}, pdfauthor={Author}]{hyperref} 
\usepackage{pythonhighlight}

\usepackage{listings}
\usepackage{markdown}

\lstnewenvironment{numbered_python}[1][]{\lstset{style=mypython,numbers=left}}{}

\definecolor{red}{rgb}{0,0,0}

\definecolor{codegreen}{rgb}{0,0.6,0}
\definecolor{codegray}{rgb}{0.5,0.5,0.5}
\definecolor{codepurple}{rgb}{0.58,0,0.82}
\definecolor{backcolour}{rgb}{0.95,0.95,0.92}
\lstdefinestyle{mystyle}{
    backgroundcolor=\color{backcolour},   
    commentstyle=\color{codegreen},
    keywordstyle=\color{magenta},
    numberstyle=\tiny\color{codegray},
    stringstyle=\color{codepurple},
    basicstyle=\ttfamily\footnotesize,
    breakatwhitespace=false,         
    breaklines=true,                 
    captionpos=b,                    
    keepspaces=true,                 
    numbers=none,                    
    numbersep=5pt,                  
    showspaces=false,                
    showstringspaces=false,
    showtabs=false,                  
    tabsize=2
}
\def\code#1{\texttt{#1}}

\newcommand{\ex}[1]{\langle #1 \rangle}

\newcommand{\fig}[1]{Fig.~\ref{#1}}

\newcommand{\beq}{\begin{eqnarray}}
\newcommand{\eeq}{\end{eqnarray}}

\newcommand{\figref}[1]{\mbox{Fig.~\ref{#1}}}

\newcommand{\secref}[1]{\mbox{Sec.~\ref{#1}}}

\renewcommand{\eqref}[1]{\mbox{Eq.~(\ref{#1})}}

\newcommand{\be}{\begin{equation}}
\newcommand{\ee}{\end{equation}}
\newcommand{\bea}{\begin{eqnarray}}
\newcommand{\eea}{\end{eqnarray}}

\newcommand*{\DeptMathAU}{Department of Mathematics, Aberystwyth University, Penglais Campus, Aberystwyth, SY23 3BZ, Wales, United Kingdom}
\newcommand*{\CEQSMQU}{Center for Engineered Quantum Systems, Dept. of Physics \& Astronomy, Macquarie University, 2109 NSW, Australia}

\begin{document}
\title{QuTiP-BoFiN: A bosonic and fermionic numerical hierarchical-equations-of-motion library with applications in light-harvesting, quantum control, and single-molecule electronics}

\author{Neill Lambert}
\email[Correspondence email address: ]{nwlambert@gmail.com}
    \affiliation{Theoretical Quantum Physics Laboratory, Cluster for Pioneering Research, RIKEN, Wakoshi, Saitama 351-0198, Japan}
    
\author{Tarun Raheja}
    \affiliation{Theoretical Quantum Physics Laboratory, Cluster for Pioneering Research, RIKEN, Wakoshi, Saitama 351-0198, Japan}

    \author{Simon Cross}
    \email[Correspondence email address: ]{hodgestar@gmail.com}
\affiliation{Theoretical Quantum Physics Laboratory, Cluster for Pioneering Research, RIKEN, Wakoshi, Saitama 351-0198, Japan}
   
\author{Paul Menczel}
\affiliation{Theoretical Quantum Physics Laboratory, Cluster for Pioneering Research, RIKEN, Wakoshi, Saitama 351-0198, Japan}

\author{Shahnawaz Ahmed}
    \affiliation{Wallenberg Centre for Quantum Technology, Department of Microtechnology and Nanoscience, Chalmers University of Technology, 412 96 Gothenburg, Sweden}

\author{Alexander Pitchford}
    \affiliation{\DeptMathAU}
    \affiliation{\CEQSMQU}

\author{Daniel Burgarth}
    \affiliation{\CEQSMQU}

\author{Franco Nori}
    \affiliation{Theoretical Quantum Physics Laboratory, Cluster for Pioneering Research,  RIKEN, Wakoshi, Saitama 351-0198, Japan}
    \affiliation{Center for Quantum Computing, RIKEN, Wakoshi, Saitama 351-0198, Japan}
\affiliation{Department of Physics, The University of Michigan, Ann Arbor, 48109-1040 Michigan, USA}
\date{\today} 

\begin{abstract}
The ``hierarchical equations of motion'' (HEOM) method is a powerful exact numerical approach to solve the dynamics and find the steady-state of a quantum system coupled to a non-Markovian and non-perturbative environment.  Originally developed in the context of physical chemistry, it has also been extended and applied to problems in solid-state physics, optics,  single-molecule electronics, and biological physics.
Here we present a numerical library in Python,  integrated with the powerful QuTiP platform, which implements the HEOM for both bosonic and fermionic environments.  We demonstrate its utility with a series of examples {\color{red} consisting of benchmarks against important known results and examples demonstrating several new results and insights}. For the bosonic case, our {\color{red} new results} include demonstrations of how to fit arbitrary spectral densities {\color{red} with different approaches}, and a study of the dynamics of energy  transfer in the Fenna-Matthews-Olson photosynthetic complex. {\color{red} For the latter, we both clarify how a suitable non-Markovian environment can protect against pure dephasing, and model recent experimental results demonstrating the suppression of electronic coherence. Importantly, we show that by combining the HEOM method with the reaction coordinate method we can observe non-trivial system-environment entanglement on timescales substantially longer than electronic coherence alone}. We also demonstrate {\color{red} new results showing} how the HEOM can be used to benchmark different strategies for dynamical decoupling of a system from its environment, and show that the Uhrig pulse-spacing scheme is less optimal than equally spaced pulses when the environment's spectral density is very broad.  For the fermionic case, we present an integrable single-impurity example, used as a benchmark of the code, and a more complex example of an impurity strongly coupled to a single vibronic mode, with applications to single-molecule electronics. 
\end{abstract}


\maketitle

\section{Introduction}

In the context of  open quantum systems \cite{Breuer02} many approaches exist to modelling the influence of an environment on a system. For example, when the interaction between system and environment is weak and Markovian (memory-less), a range of perturbative methods are available, primarily in the form of generalized Lindblad master equations \cite{Lindblad76,gorini}, or non-secular master equations \cite{Breuer02,Strasberg_2016}. However, beyond these perturbative limits one must start to treat the dynamics of the environment on the same level as the system, which in many cases is a challenging task. 

Discretization of a continuum environment is one powerful approach to this non-perturbative limit \cite{Chin2012,chin2013role,iles2014environmental,Lambert2019,TEMPO,ACE}, wherein the predominant collective degrees of freedom are identified and included in the full simulation in a numerically efficient way.  The ``hierarchical equations of motion'' (HEOM) method does precisely this, albeit indirectly \cite{Tanimura_1989,ishizaki2005quantum,Akihito09,ishizakireview,Tanimura2020}. It is based on the construction of a hierarchy of coupled equations resulting from taking repeated time-derivatives of an influence functional, under the assumption that the bath correlation functions take an exponential form. Such an exponential form can be acquired analytically for certain spectral densities (with Matsubara or Pad\'e decompositions \cite{BetterPade,shi2009efficient}) or numerically with fitting \cite{tannor-meier}. Once the set of equations is obtained, they can be numerically solved to give the system dynamics, the system steady-state, and certain bath properties. The limitations of the HEOM arise in the truncation of the set of equations (which in the case of a bosonic environment is infinite) and the truncation of the  exponential decomposition of the bath correlation functions. The latter limits the accuracy to represent the real problem and can lead to unphysical results if poor decompositions are made. 

In this paper we introduce an open-source library for modelling these equations of motion for both bosonic and fermionic baths as an integrated part of our larger toolbox for modelling open quantum systems in Python, QuTiP \cite{Johansson12, Johansson13,Shammah18,boxi}.  We describe the bosonic (\secref{sec:bos}) and fermionic (\secref{sec:ferm}) forms separately. For both we start with basic definitions, and then provide a range of examples, some of which reproduce {\color{red} important} known results, and {\color{red} three of which} give novel insights gained with this tool for this article:
\begin{itemize}
    \item In \secref{sec:spin_boson_model} we apply the HEOM to the ubiquitous spin-boson model. We show results for the Drude-Lorentz spectral density, and the under-damped Brownian motion spectral density, and show convergence trends for the Matsubara, Pad\'e, and fitting  approaches to decomposing the bath correlation function. We also compare the results to a standard Markovian method in the form of the Bloch-Redfield solver in QuTiP.
    \item In \secref{sec:fmo} we demonstrate how to model multiple baths by reproducing the seminal results of \cite{Akihito09} for the dynamics of energy transport through the Fenna-Matthews-Olson pigment protein complex, and showing that the non-Markovian nature of the environment preserves electronic coherence as compared to a standard Bloch-Redfield approach. {\color{red} We also take the recent experimental results of  \cite{Duan8493} and, using a combination of the HEOM and the reaction-coordinate method, show that system-environment entanglement oscillations persist on longer time-scales than electronic coherence in a strongly-damped regime.}
    \item In \secref{sec:therm} we illustrate how the auxiliary density operators in the HEOM can be used to obtain information about the environment by reproducing results from \cite{KatoJChemPhys2015} regarding non-equilibrium heat flow into an environment.
    \item In \secref{sec:dd} we demonstrate how capturing the non-Markovian nature of the environment can be important for quantum control of quantum systems by examining the interplay between dynamical decoupling schemes and environment properties. Importantly, we both demonstrate how the HEOM can be used in regimes where standard analytical results fail (finite control pulse length), and benchmark two pulse-spacing schemes against each other (equally spaced versus Uhrig's optimal strategy).
    \item In \secref{sec:fitting} we show how to go beyond the standard choices of spectral density with the HEOM using different fitting techniques. We explicitly demonstrate how to capture an Ohmic spectral density with exponential cut-off using either spectral-density fitting or correlation function fitting techniques, {\color{red} and benchmark them against each other using the pure-dephasing model. }
    \item In \secref{sec:resonant} we benchmark the fermionic solver against an analytical result for the current through a single resonant level coupled to two fermionic baths.
    \item In \secref{sec:resonanceandboson}, for the fermionic solver, we reproduce the results in \cite{schinabeck2}  showing the steady state current through a resonant level coupled both to two fermionic baths and to a single vibronic  mode.
\end{itemize}

Each example is associated with a Jupyter Notebook provided on GitHub \cite{examindex}, which throughout this paper are referred to via the numbered index in their name, 1a, 1b, etc.

\subsection{Comparison to existing packages}
{\color{red}
Other open-source implementations of the HEOM exist, a summary of which, including the one presented in this work, is given in table~\ref{sum}. It is difficult to be exhaustive on the capabilities and what constitutes useful levels of documentation or testing for all these packages. Hence, for documentation, a "yes" was given if the available documentation seemed sufficient for a third-party to install and run the software on their own problems, without having to read the source code. A "yes" was given for tests if the software included a  suite of tests that cover the main features of the implementation and checks the resulting outputs. 
QuTiP's implementation is the only one that meets this standard. 

In summary, Tanimura provided an early Fortran implementation on his website \cite{TanimuraCode}. PHI \cite{Strumpfer2012,SchultenCode} was a later implementation that provided multithreaded CPU support. The Nanohub GPU HEOM \cite{kreisbeck} (unrelated to the Tsuchimoto GPU-HEOM \cite{Tsuchimoto2015}) provided GPU support but is closed source. The Tsuchimoto GPU-HEOM \cite{Tsuchimoto2015} (also from the Tanimura group) implements a custom matrix exponential solver to reduce memory overhead. HEOM-QUICK \cite{HEOMQUICK} is the only other implementation to support fermionic baths, and provides much flexibility via editing functions in the Fortran code. DM-HEOM \cite{HEOMDM,Noack,noack2}  supports distribution across multiple nodes, allowing the simulation of systems where the memory requirements of the hierarchy exceed that available on a single node. PyHEOM \cite{ikeda2020}  provides explicit support for a new decomposition method for capturing poles in the bath spectral densities.


\begin{table*}[!ht]
    \centering
    \begin{tabular}{|p{30mm}|l|l|l|p{30mm}|l|}
    \hline
        Name & Docs & Tests & T-D Hamiltonian & Baths & Compute \\ \hline
        PHI \cite{Strumpfer2012,SchultenCode} & Yes & No & No & Bosonic DL & Single node, multithreaded CPU \\ \hline
        Nano GPU HEOM \cite{kreisbeck}& Yes & No & No & Bosonic DL & Single node, multithreaded-CPU or GPU \\ \hline
        GPU-HEOM \cite{Tsuchimoto2015}  & No & No & No & Bosonic DL & Single node GPU \\ \hline
        HEOM-QUICK \cite{HEOMQUICK}.  & Yes & No & Yes & Fermionic Lor  & Single node, multithreaded-CPU \\ \hline
        DM-HEOM \cite{HEOMDM,Noack,noack2} & No & No & No & Bosonic DL & Multiple node, multithreaded-CPU and GPU \\ \hline
        PyHEOM \cite{ikeda2020} & No & No & No & Bosonic, DL, UD & Single node, single-threaded CPU \\ \hline
        QuTiP-BoFiN \newline (this work) & Yes & Yes & Yes & Bosonic DL, UD, fit, \newline \& Fermionic Lor, fit & Single node, multithreaded-CPU \\ \hline
    \end{tabular}
    \label{sum}
   \caption{{\color{red}  A table comparing available HEOM implementations. The terminology used is as follows.
Name:
  Abbreviated name of the implementation.
Docs:
  ``Yes'' if there seemed to be sufficient documentation for a third-party to install and run the software on their own problems, without having to read the source code.
Tests:
  ``Yes'' if the software included a suite of tests that cover the main features of the implementation and checks the resulting outputs.
Time-dependent Hamiltonians:
  ``Yes'' if time-dependent Hamiltonians are supported. 
Baths:
  A description of which types of baths are supported. DL refers to Drude-Lorentz, UD refers to Under-damped Brownian motion, ``Lor'' refers to Lorentzian (for fermionic baths), and ``fit'' refers to fitting via generic decomposition of correlation functions.
Compute:
  A description of what computational paradigms are supported. We distinguish between GPU, CPU, multithread, and single-versus-multiple
  computational nodes.} }
\end{table*}


Our implementation is complementary to the above packages, in the sense that it provides flexible implementations of both bosonic and fermionic cases, and allows for both pre-defined commonly used spectral densities and arbitrary user-based input (that can be combined with custom fitting of bath correlation functions or spectra, as required). In addition, integration with the QuTiP library will ensure its continuous maintenance and improvement by an established team of developers~\cite{qutipdevs}.
}
\section{Bosonic Environments}\label{sec:bos}

\subsection{Basic Definitions}
When considering the influence of an environment on a quantum system, a typical starting point is to consider the environment as a bath of linear harmonic oscillators.  In some physical situations this is justified by the actual dominant degrees of freedom being harmonic in nature.  However, it is also a very pragmatic and often phenomenological assumption, as, when combined with the assumption that the bath is initially in a Gaussian (e.g., thermal) state, evaluating the influence of the bath on the system is much easier. In this work we do not consider non-linear baths (e.g., spin-baths), though it has been shown that such environments can be captured with the HEOM to some degree \cite{CaoSpin1,CaoSpin2}.

In the standard second-quantized Hamiltonian formalism, we can write the interaction of a single bosonic environment with an arbitrary system in the following way,
\beq
H = H_{\mathrm{S}}(t) + \sum_k \omega_k a_k^{\dagger}a_k + Q \sum_k g_k \left(a_k + a_k^{\dagger}\right).
\label{H}
\eeq
where $H_S(t)$ is the free Hamiltonian of the system, which can be time-dependent, $H_{\mathrm{B}} = \sum_k \omega_k a_k^{\dagger}a_k$ is the free Hamiltonian of the bath, $Q$ is the system operator which couples to the bath, and $g_k$ is the coupling strength between the system and mode $k$ in the bath. Note that this, and the following, is easily generalized to multiple baths with different system coupling operators.  For convenience, we define 
\beq 
X = \sum_k g_k (a_k + a_k^{\dagger}).
\eeq
Note that in the above, and throughout most of this work, we set $\hbar = k_\mathrm{B} = 1$ (some later examples are presented in different units).

The state of the system at time $t$ has an exact solution in the form of a time-ordered influence functional, which only depends on the free bath correlation functions \cite{Tanimura_1989,ishizaki2005quantum,ma2012entanglement,Breuer02},
\begin{widetext}
\beq\label{influence}
\bar{\rho}_{\mathrm{S}}(t) = \mathcal{T}\exp \left\{- \int_0^t dt_2 \int_0^{t_2}dt_1 \bar Q(t_2)^{\times} \left[C_R(t_2-t_1) \bar Q(t_1)^{\times} + i C_I(t_2-t_1) \bar Q(t_1)^\circ \right] \right\}\bar{\rho}_{\mathrm{S}}(0)
\eeq
\end{widetext}
Here $\bar{\rho}_S(t)$ is the density matrix of the system in the interaction picture with respect to $H_S(t) + H_B$ (indicated by the bar) at time $t$, and there is an implicit assumption that at $t=0$ the system and the environment are in a product state $\rho(t=0) = \rho_S(t=0)\otimes \rho_B$ (the sub-index $B$ refers to the bath or environment).  The environment is an initially thermal equilibrium state 
\beq
\rho_B = \frac{e^{-\beta H_B}}{Z}
\eeq 
with temperature $T=1/\beta$ and $Z=\mathrm{Tr}[e^{-\beta H_B}]$.  Importantly, the system operators $Q$ that couple to the environment  act to the right as superoperators in \eqref{influence}, with 
\beq \label{xo}
\bar Q(t)^{\times}=[\bar Q(t),\centerdot],\quad \mathrm{and} \quad \bar Q(t)^\circ=\left\{\bar Q(t),\centerdot\right\}.  
\eeq

The derivation of this exact result, \eqref{influence}, relies crucially on the Gaussian nature of the free environment, which allows the properties of the system to only depend on the second order correlation functions of the environment.   These correlation functions can be expressed as,

\beq\label{ct}
C(\tau) &=& \ex{\bar{X}(t+\tau)\bar{X}(t)} \nonumber\\
&=& \mathrm{Tr}\left[\sum_k g_k(a_ke^{-i\omega_k \tau} + a_k^{\dagger}e^{i\omega_k \tau})\sum_{k'} g_{k'}(a_{k'} + a_{k'}^{\dagger})\rho_B\right]\nonumber\\
&=& \sum_kg_k^2\left[n(\omega_k,\beta)e^{i\omega_k \tau} +\left\{1+n(\omega_k,\beta)\right\}e^{-i\omega_k \tau}\right]\nonumber\\
&=& \int_0^{\infty} d\omega \frac{J(\omega)}{\pi}\left[ n(\omega,\beta)e^{i\omega \tau} +  \left\{1+n(\omega,\beta)\right\}e^{-i\omega \tau} \right]\nonumber\\
&=& \int_0^{\infty} d\omega \frac{J(\omega)}{\pi} \left[\coth(\beta\omega/2) \cos(\omega \tau) - i \sin(\omega \tau) \right]
\eeq
where $n(\omega, \beta) = 1/[\exp(\beta  \omega) -1]$, and in moving from the discrete to continuum limit one defines 
\beq
J(\omega) = \pi \sum_k |g_k|^2 \,\delta(\omega-\omega_k).
\eeq
Returning to \eqref{influence}, which is formally exact but still a time-ordered exponential that is difficult to directly solve, if we assume that an appropriate choice of $J(\omega)$ gives correlation functions 
\beq
C(t) = C_R(t) + i C_I(t)
\eeq
with real and imaginary parts that can be decomposed as,
\beq \label{exponens}
C_R(t) &=& \sum_{k=1}^{N_R} c_k^R e^{-\gamma_k^R t}\\
C_I(t) &=& \sum_{k=1}^{N_I} c_k^I e^{-\gamma_k^I t}
\eeq
where $c_k^j$ and $\gamma_k^j$ themselves can be real or complex, one can formally take repeated time derivatives of \eqref{influence} to arrive at an infinite set of coupled first-order differential equations.  A concise description of this procedure can be found in the appendix of [\onlinecite{ma2012entanglement}]  for a particular case (see also [\onlinecite{hartle}] for an alternative approach, and [\onlinecite{Tanimura_1989}] for the original derivation using path integrals), but for the most general decomposition one ultimately finds the following set of coupled differential equations \cite{fruchtman2016perturbative},
\beq\label{heomB}
\dot{\rho}^{n}(t)&=&\left(\mathcal{L} - \sum_{j=R, I} \sum_{k=1}^{N_{j}} n_{j k} \gamma_{k}^{j}\right) \rho^{n}(t) \nonumber\\
&-&i \sum_{k=1}^{N_{R}} c_{k}^{R} n_{Rk} Q^{\times} \rho^{n_{R k}^{-}}(t)
+\sum_{k=1}^{N_{I}} c_{k}^{I} n_{Ik} Q^{\circ} \rho^{n_{I k}^{-}}(t) \nonumber\\
&-&i \sum_{j=R, I} \sum_{k=1}^{N_{j}} Q^{\times} \rho^{n_{ j k}^{+}}(t)
\eeq
where $n = (n_{R1},n_{R2},..n_{RN_R}, n_{I1},n_{I2},...,n_{IN_I})$ is a multi-index of integers $n_{jk} \in \{0..N_c\}$, and $N_c$ is a cut-off parameter chosen for convergence. The state labelled by $(0,...,0)$ describes the system density matrix. Operators with non-zero indices are referred to as auxiliary density operators (ADOs), but are not physical density operators in the normal sense. They correspond to terms collected for different exponents of the correlation functions that appear from the application of the chain-rule to the time-derivatives of the time-ordered integral. Terms like $\rho^{n_{ j k}^{\pm}}$ in \eqref{heomB} refer to an ADO with the index $n_{jk}$ raised or lowered by one. The benefit of \eqref{heomB} is that we can truncate and solve the finite set of coupled differential equations with a range of efficient numerical methods.

The Liouvillian describes the local (potentially time-dependent) system Hamiltonian 
\beq
\mathcal{L}=- i H_{S}^{\times},
\eeq
where we use the same notation for the commutator as in \eqref{xo}.  In some cases this can be augmented with additional Lindblad terms, as described later.

If there are terms in the decomposition of the correlation function with equal frequencies $\gamma_k^R = \gamma_k^I$ they can be combined into a single index, for a gain in numerical efficiency \cite{fruchtman2016perturbative}.  This is done automatically by our code itself.  Due to the linearity of the interaction with the bath, additional environments can be included by simply extending the list of bath correlation functions (i.e., adding a bath label $K$ to \eqref{exponens}, and different system coupling operators to different baths is allowed for by labelling the indices and the coupling operator $Q_K$ appropriately.

\subsection{Code Functionality}

Our Python-based library relies on, and is integrated with, the popular Quantum Toolbox in Python (QuTiP) \cite{Johansson12, Johansson13}, and hence uses the ubiquitous \code{Qobj} data types from that library to represent states, operators and superoperators.
Generally speaking, one sets up the problem to be solved by defining the system Hamiltonian (which can be time-dependent), the bath properties using either pre-defined bath class objects, or through lists defining the correlation function decomposition given in \eqref{exponens} and a generic bath class, and  a system coupling operator for each bath.

The \code{HEOMSolver} class constructs the right-hand-side of \eqref{heomB}, {\color{red} taking as input one or more baths (which can be defined in multiple ways), the system Hamiltonian, and the truncation parameters. For example, for a bosonic bath using the predefined Drude-Lorentz spectral density with $N_k$ Matsubara terms:
}
\begin{lstlisting}[language=Python, caption=Defining the hierarchy of equations]
bath = DrudeLorentzBath(Q, lam=lam, gamma=gamma, T=T, Nk=Nk, tag="DLBath")
HEOM_dlbath = HEOMSolver(Hsys, bath, NC)
\end{lstlisting}

{\color{red}

Here \code{Q} is the system-bath coupling operator, and the other parameters determine the properties of the Drude-Lorentz spectral density \eqref{JD}. In addition, one can define a bosonic bath using the same spectral density with a Pade decomposition [\code{DrudeLorentzPadeBath()}], the underdamped spectral density of \eqref{JU} [\code{UnderDampedBath()}], or a generic one using the real and imaginary coefficients of \eqref{fit} [\code{BosonicBath()}]. 

This \code{HEOMSolver} object can then be used to solve the coupled set of ordinary differential equations using standard libraries, or calculate the steady-state. For the former purpose, the class contains a \code{run()} function which takes an initial condition and a set of time-steps, and returns a results object with both the system density matrix and the auxiliary-density-operators (ADOs) for the bath at each time-step:
}
\begin{lstlisting}[language=Python, caption=Solving the time evolution]
result_dlbath = HEOM_dlbath.run(rho0, tlist,ado_return=True)
\end{lstlisting}
{\color{red}
The state of the system is returned for each time step in \code{result\_dlbath.states}, while, if \code{ado\_return=True} is used, then the full state of the auxiliary density operators is returned as a list as well, in \code{result\_dlbath.ado\_states}. These can then be used to obtain certain bath properties. Tags can be employed to delineate different baths, and levels can be used to extract different tiers. 

For example, for the example defined above, \code{ado\_states.filter(level=1, tags=["bathDL"])} returns a list of labels 
identifying the ADOs that are tagged with \code{"bathDL"} and are at truncation tier $1$ (i.e., only have $\sum_{j=R,I}\sum_{k=1}^{N_j} n_{jk} = 1$ for the given bath). Using \code{result\_dlbath.extract()}, these labels can then be used to extract specific ADOs, see Example 3 and its associated notebook for a practical example.  }

For computing the steady-state solution, we provide two direct methods, one of which takes advantage of Intel's MKL library if installed alongside QuTiP.  Full details of the functionality can be seen in the accompanying Jupyter notebooks (provided in [\onlinecite{examindex}]) and documentation \cite{qutipweb3}.

\subsection{Example 1: Spin-boson model} \label{sec:spin_boson_model}

The archetypical test system for studying open quantum systems is a two-level system (i.e., a spin, or qubit, but we use the terminology two-level system throughtout this work). In the context of physical chemistry, it is used as a model of electronic energy transport between two nearby molecules (i.e., a dimer model). Referring back to \eqref{H}, in this example we describe the system Hamiltonian using the standard Pauli operators to represent the two-level system
\beq
H_{\mathrm{S}} &=& \frac{\epsilon}{2}\sigma_z + \frac{\Delta}{2}\sigma_x 
\label{HS}
\eeq
and the coupling operator 
\beq
Q&=& \sigma_z
\label{Q}
\eeq

Ultimately the choice of spectral density, $J(\omega)$, depends on the properties of the environment one is considering.  A useful choice is that of an Ohmic (linear) frequency dependence with a Lorentzian cut-off.  This is usually split into two types, the over-damped Drude-Lorentz form
\beq
J_D(\omega) = \frac{2\lambda \gamma \omega}{(\gamma^2 + \omega^2)}
\label{JD}
\eeq
and the under-damped Brownian motion form,
\beq
J_U(\omega) = \frac{\alpha^2 \Gamma \omega}{[(\omega_c^2 - \omega^2)^2 + \Gamma^2 \omega^2]}.\label{JU}
\eeq
While superficially similar, the properties of the corresponding correlation functions ($C_D(t)$ and $C_U(t)$) are different in several important ways.  Firstly, as we will show below, with a Matsubara decomposition $C_D(t)$ has only overdamped exponents, while $C_U(t)$ can have oscillatory parts.  Secondly, $J_D$ has a $1/\omega$ high-frequency tail, while $J_U$ converges much more quickly.  This leads to two unique features in $C_D(t)$ at $t=0$:  a non-zero imaginary part (which is arguably unphysical \cite{ishizakiDLsucks}),  and a divergent real part.  

For example, the Matsubara decomposition of the Drude-Lorentz spectral density is given by:

\begin{equation}\label{ctd}
C(t)=\sum_{k=0}^{\infty} c_k e^{-\nu_k t}
\end{equation}

\begin{equation}
    \nu_k = \begin{cases}
               \gamma               & k = 0\\
               {2 \pi k} / {\beta }  & k \geq 1\\
           \end{cases}
\end{equation}

\begin{equation}
    c_k = \begin{cases}
               \lambda \gamma [\cot(\beta \gamma / 2) - i]             & k = 0\\
               4 \lambda \gamma \nu_k / (\nu_k^2 - \gamma^2)\beta    & k \geq 1.\\
           \end{cases}
\end{equation}
 The divergent real part at $t=0$ can be captured in the HEOM formalism by treating exponentials above a certain cut-off, $N_k$, as delta functions \cite{ishizaki2005quantum}.  Since $\nu_k=2 \pi k/\beta $, if $1/\nu_k$ is much smaller than other important time-scales, we can approximate,  $ e^{-\nu_k t} \approx \delta(t)/\nu_k$, and 
 \beq 
 C(t)\approx \sum_{k=0}^{N_k} c_k e^{-\nu_k t} +
 \sum_{k=N_k+1}^{\infty} \frac{c_k}{\nu_k} \delta(t).
 \eeq
It is convenient to calculate the whole sum 
\beq
\sum_{k=0}^{\infty} \frac{c_k}{\nu_k} =  \frac{2 \lambda }{ \beta \gamma} - i\lambda,
\eeq and subtract off the contribution from the finite $N_k$ Matsubara terms that are kept in the hierarchy, and treat the residual as a prefactor multiplying a delta function. It is then possible to show  \cite{ishizaki2005quantum} that the delta-function contribution to the correlation functions can be described by Lindbladian terms in the HEOM; and thus the equation of motion in \eqref{heomB} is  modified so that the Liouvillian evolution includes these Lindblad terms
\beq
\mathcal{L}=- i H_{S}^{\times} + \Gamma_T [2Q\rho Q^{\dagger} - Q^{\dagger}Q\rho - \rho Q^{\dagger}Q],
\eeq
where 
\beq
\Gamma_T =   \frac{2 \lambda}{\beta \gamma} - i\lambda  - \sum_{k=0}^{N_k} \frac{c_k}{\nu_k}.\eeq  This treatment is sometimes called the ``Tanimura-terminator'', and an example of how to include it is presented in the accompanying example notebook 1a\cite{examindex}.


    

As described earlier, the QuTiP-BoFiN package~\cite{qutipweb1,qutipweb2} will automatically determine that real and imaginary exponents are close, and if the coupling operator for both exponents are equal, it will combine them into a single common exponent.  Note that for degenerate exponents in the real or imaginary terms alone the code will not automatically combine terms.
\subsubsection{Two-level system coupled to a Drude-Lorentz bath}

\begin{figure}[]

\includegraphics[width = \columnwidth]{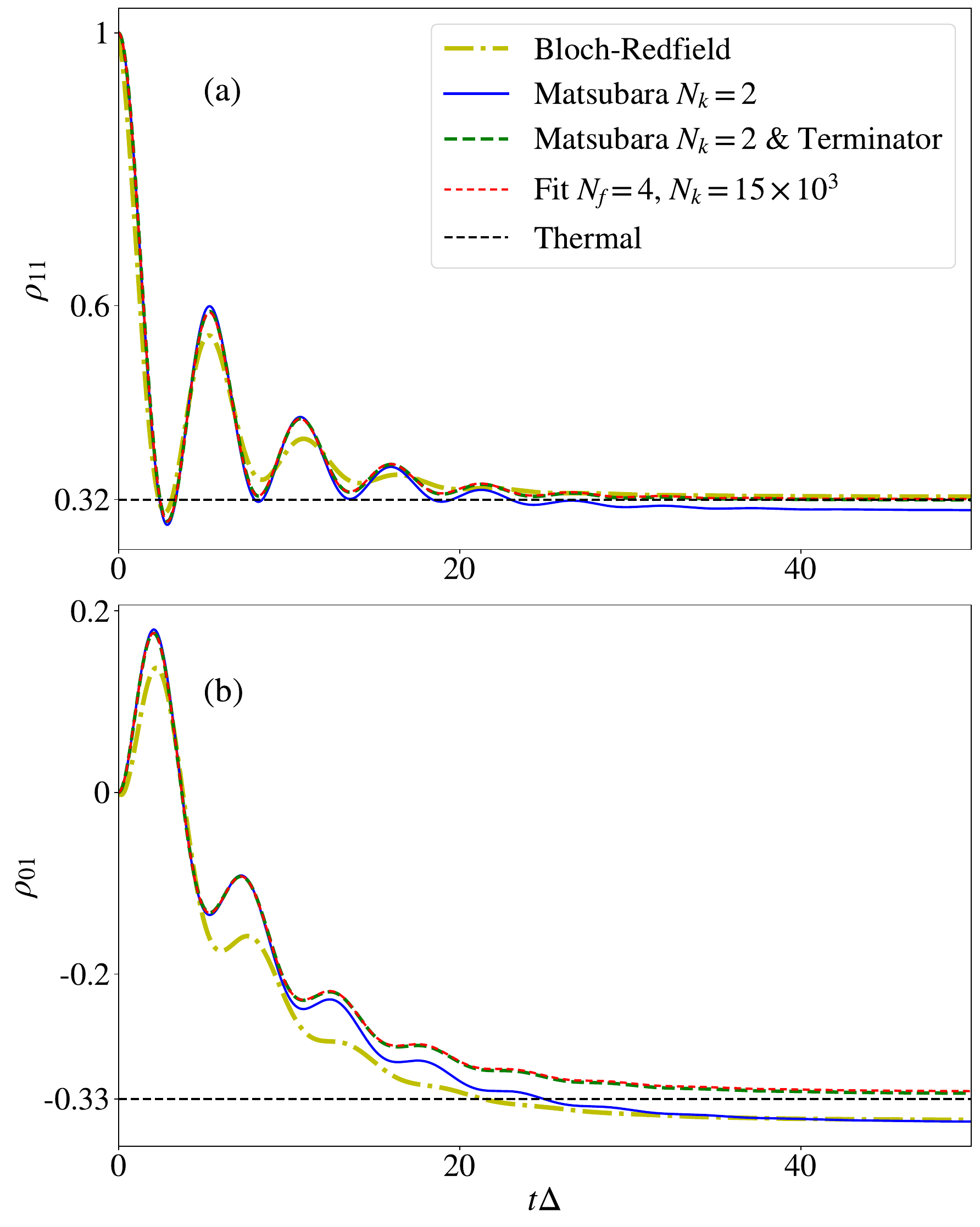}
\caption{HEOM results for the excited state probability $\rho_{11}$ and the coherence $\rho_{01}$ versus time for a two-level system coupled to a bosonic environment described by a Drude-Lorentz spectral density. The system parameter used here is $\epsilon = 0.5 \Delta$, and the bath parameters are $\lambda = 0.1 \Delta$, $\gamma=0.5 \Delta$, and $T=0.5 \Delta$.  We compare a solution of the Bloch-Redfield master equation to the Matsubara decomposition with $N_k=2$, the same decomposition with the Tanimura terminator, and the case where only the primary (Drude) exponent ($k=0$) is kept in \eqref{ctd}, and a large number ($N_k=15\times 10^3$) of Matsubara terms are summed and fit with $N_f=4$ exponents.  The horizontal dashed black line shows the thermal (steady) state result obtained from a reaction-coordinate approach. The code for generating this figure can be found in example notebook 1a \cite{examindex}.}\label{fig1}
\end{figure}

In \figref{fig1} we show a simple example of the dynamics of a two-level system coupled to a Drude-Lorentz bath. We compare several optional approaches. This includes an explicit truncation of the Matsubara decomposition at $N_k=2$, the same truncation with the Tanimura terminator,
and an example where a large number of Matsubara terms ($N_k = 15 \times 10^3$) are summed up and numerically fit using the  fitting function $C_F = \sum_{i=1}^{N_f} a_i e^{b_i t} $ with $N_f = 4$ exponents. We also show the solution from the standard Bloch-Redfield solver in QuTiP (i.e., a generalized Born-Markov-Secular master equation).  In the accompanying example notebook 1a \cite{examindex} we also present the results of a Pad\'e decomposition of the correlation functions, but do not show it in the figure (it essentially converges faster than the Matsubara case alone as it gives the same result as the Matsubara plus terminator decomposition, but without the need of an equivalent terminator).

We can see from \figref{fig1} that, for this choice of parameters, either the Tanimura-terminator, the Pad\'e decomposition, or the fitting approach is needed to reach the correct steady-state (alternatively more discrete terms must be included in the Matsubara summation).  The black dashed lines show the steady-state obtained using a reaction-coordinate approach \cite{iles2014environmental}.  This trend can be confirmed with a pure-dephasing analytical result, which is provided in example notebook 1e \cite{examindex}, but not explicitly shown here.
\subsubsection{Two-level system very strongly coupled to a Drude-Lorentz bath}
\begin{figure}[]
\includegraphics[width = \columnwidth]{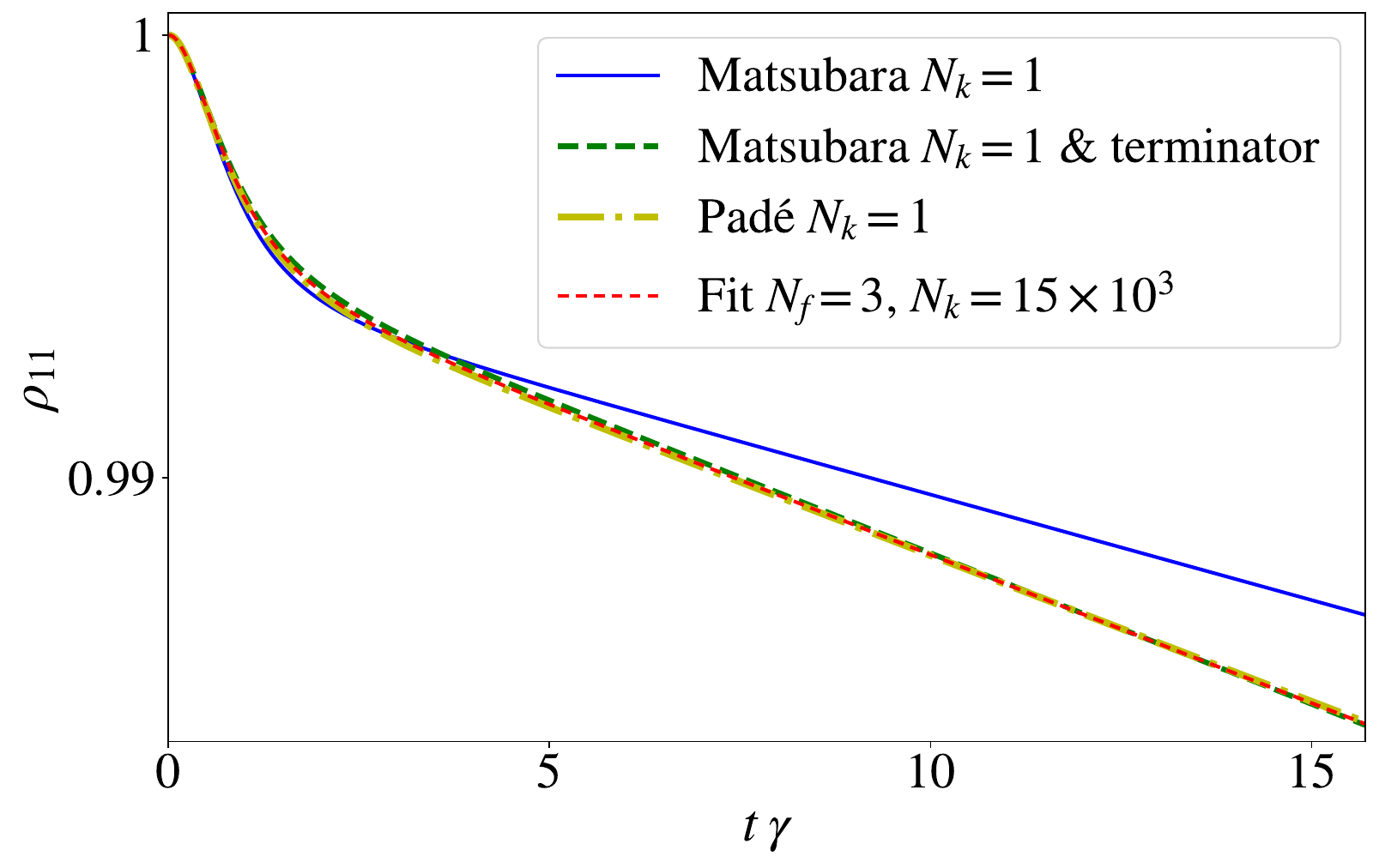}
\caption{HEOM results for the excited state probability versus time for a two-level system coupled to a bosonic environment with a  Drude-Lorentz spectral density in the strong coupling regime. Here we use the system parameter $\epsilon = 0.0$, $\Delta = 0.2 \gamma$,  and the bath parameters $\lambda = 2.5 \gamma$,  and $T=\gamma$.  We show the case of a single Matsubara exponent, the same with the Tanimura terminator, the Pad\'e decomposition also with a single (non-Drude) exponent, and again a fit of $N_k=15\times 10^3$ Matsubara terms, this time with $N_f =3$ exponents. The code for generating this figure can be found in example notebook 1b \cite{examindex}.}\label{fig2}
\end{figure}

Moving to a more challenging parameter regime (see also example notebook 1b \cite{examindex}), \figref{fig2} shows the situation of a very strong coupling between system and environment (as used as a benchmark in [\onlinecite{shi2009efficient}]).   We compare  the results obtained with a single Matsubara term, the same with the terminator included, a single Pad\'e decomposition term, and a case where we sum up a large number of Matsubara terms and fit them with auxiliary exponents.
Here the Bloch-Redfield approach fails completely, so it is not explicitly shown.  As described in [\onlinecite{shi2009efficient}], in this case the terminator result converges very quickly, while many discrete Matsubara terms would be needed without it. The Pad\'e result also converges faster than the single-term Matsubara result, as does the result using a fit, but with a higher numerical overhead.  We expect a more sophisticated fitting algorithm may provide better results (as evidenced by the better performance of the Pad\'e decomposition with less exponents).

\subsubsection{Two-level system coupled to an underdamped bath}

Moving to the underdamped case, the underdamped spectral density gives correlation functions (again using the Matsubara decomposition) separated explicitly into real and imaginary parts of the form:
\begin{equation} \label{cU}
    c_k^R = \begin{cases}
               \alpha^2 \coth(\beta( \Omega + i\Gamma/2)/2)/4\Omega & k = 0\\
               \alpha^2 \coth(\beta( \Omega - i\Gamma/2)/2)/4\Omega & k = 0\\
 \frac{-2\alpha^2\Gamma\epsilon_k }{\beta[(\Omega + i\Gamma/2)^2 + \epsilon_k^2)][(\Omega - i\Gamma/2)^2 + \epsilon_k^2]}      & k \geq 1\\
           \end{cases}
\end{equation}

\begin{equation}
    \nu_k^R = \begin{cases}
               -i\Omega  + \Gamma/2 & k=0 \\
               i\Omega  +\Gamma/2   & k = 0\\
               {2 \pi k} / {\beta }  & k \geq 1\\
           \end{cases}
\end{equation}

\begin{equation}
    c_k^I = \begin{cases}
               i\alpha^2 /4\Omega & k = 0\\
                -i\alpha^2 /4\Omega & k = 0\\
           \end{cases}
\end{equation}

\begin{equation}
    \nu_k^I = \begin{cases}
               i\Omega  + \Gamma/2 &k=0\\ 
               -i\Omega  + \Gamma/2  & k = 0\\
           \end{cases}
\end{equation}

Where $\Omega = \sqrt{\Omega_c^2 - (\Gamma/2)^2}$.
Note that here we slightly abuse notation to include the pair of complex conjugate terms with one index ($k=0$). These are presented as separate terms in the real and imaginary decomposition here, but the package will again combine real and imaginary parts of the $k=0$ components with equal exponents, such that there are two total effective exponents instead of four.

\begin{figure}[h]
\includegraphics[width = \columnwidth]{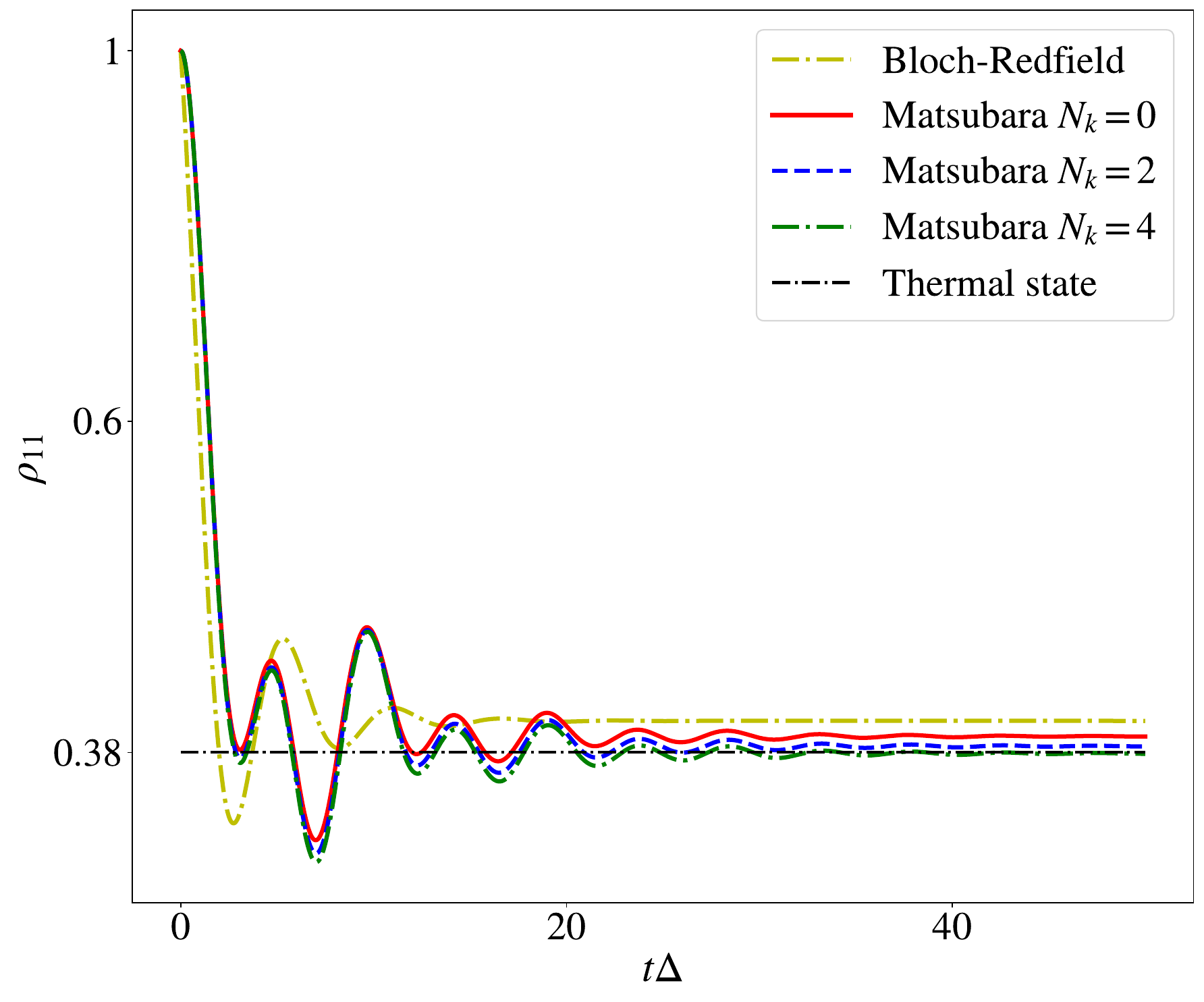}
\caption{HEOM result for the excited state probability versus time for a two-level system coupled to a under-damped spectral density with system parameter $\epsilon = 0.5\Delta$, and {\color{red} bath parameters $\alpha = 0.5 \Delta^{3/2}$, $\Gamma=0.5 \Delta$, $\omega_c = \Delta$ and $T=0.05\Delta$. Here we show how the result converges as we increase the number of Matsubara terms, and compare to the solution from the Bloch-Redfield master equation solver (which fails drastically). The black dashed line represents the thermal steady-state predicted by the reaction coordinate method.} The code for generating this figure can be found in example notebook 1c \cite{examindex}.}\label{fig3}
\end{figure}

In \figref{fig3} (see also example notebook 1c \cite{examindex}) we show a typical result for a strongly coupled, narrow spectral density. We see a large deviation between the Born-Markov-Secular  Bloch-Redfield result and the HEOM result, {\color{red} and a convergence of the HEOM result with around four Matsubara terms. In particular, we observe that the Matsubara terms, for narrow spectral densities like this one, largely contribute to correcting detailed balance and arriving at the correct steady state. We benchmark this effect by plotting the steady-state using the reaction coordinate (RC) method \cite{iles2014environmental}, which shows how as we add more Matsubara terms to the HEOM approach we tend towards the  steady-state predicted by the RC method  (see [\onlinecite{Lambert2019}] for a more detailed analysis of this effect, and results that include fitting of Matsubara terms to reach $T=0$).}

\subsection{Example 2: FMO-complex} \label{sec:fmo}

The HEOM method has a long history of being applied to problems in physical chemistry, particularly in the study of transport of electronic excitations through light-harvesting complexes \cite{Akihito09,Lambert13,Chen2015,Li2012}.  The typical parameters of such systems  sit in a regime where it is hard to justify any particular perturbative approximation, as the electronic dipole coupling between chromophores and the dominant environmental influence are, in many cases, of the same order.  One of the most studied examples of this type of problem is that of the Fenna-Mathews-Olson (FMO) complex.  It serves us well here as a way to show how to use our package for a system with several internal levels coupling to independent environments.

One again starts with a Hamiltonian description of the system. We model a single excitation moving through the FMO complex with a seven-site model
\beq
H_S = \sum_i \epsilon_i \ket{E_i}\bra{E_i} + \sum_{i \neq j} J_{i,j} \ket{E_i}\bra{E_j}
\eeq
where the on-site energy of site $i$ is given by $\epsilon_i$ and the dipole coupling between bacteriochlorophyls (BChl) $i$ and $j$ is given by $J_{i,j}$.

Again we treat the environment as a continuum of harmonic oscillators, and assume that each BChl is coupled to an independent but identical bath of such oscillators. This represents how the harmonic nuclear motion of the molecules leads to changes in the electronic energies.  The coupling operator for the $i$-th bath is thus $Q_i =  \ket{E_i}\bra{E_i}$, and we assume each independent bath is described by a Drude-Lorentz spectral density.
A separate bath class object can be created for each bath, associated with each system coupling operator, using whatever bath-decomposition scheme one requires. An explicit example of this is given in example notebook 2 \cite{examindex}.  


%

\begin{figure}[]
\includegraphics[width = \columnwidth]{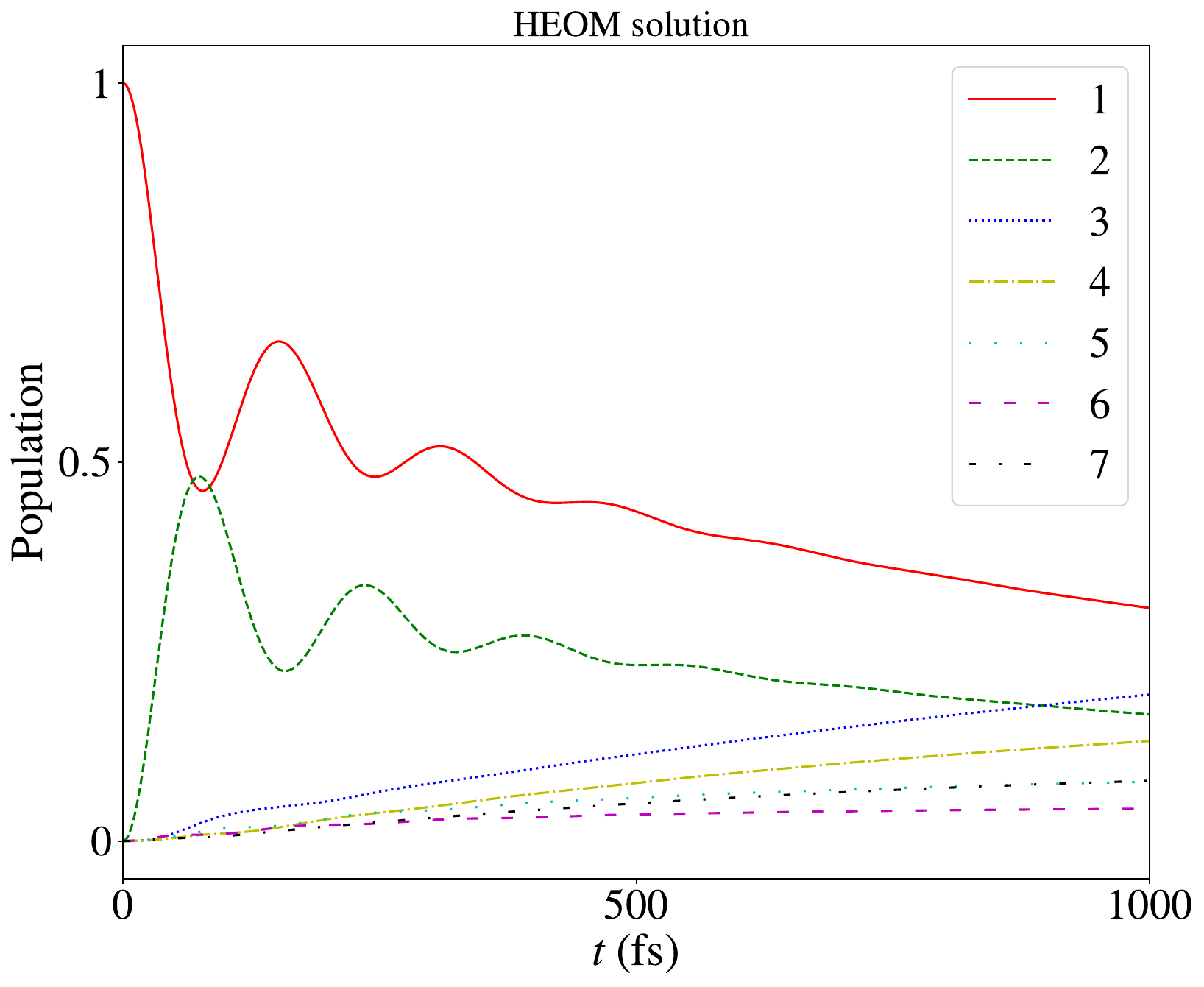}
\caption{HEOM results for the site-populations of the FMO complex versus time. The FMO complex consists of seven sites coupled to seven identical but independent baths each described by a Drude Lorentz spectral density with parameters $\lambda = 35$ cm$^{-1}$, $\gamma^{-1}$ = $166$ fs, $T=300$ K. For the FMO Hamiltonian we use the data employed in \cite{Akihito09}.  In comparison to \figref{fig5} we see that the exact solution predicts much longer coherent oscillations than the equivalent weak-coupling solution. The code for generating this figure can be found in example notebook 2 \cite{examindex}.}\label{fig4}
\end{figure}

We provide an example of simulation results in \figref{fig4} showing how, with the initial condition of an excitation localised on site 1, the populations evolve up to $1$ ps.  These results identically recreate those seen in [\onlinecite{Akihito09}]. 

\begin{figure}[]
\includegraphics[width = \columnwidth]{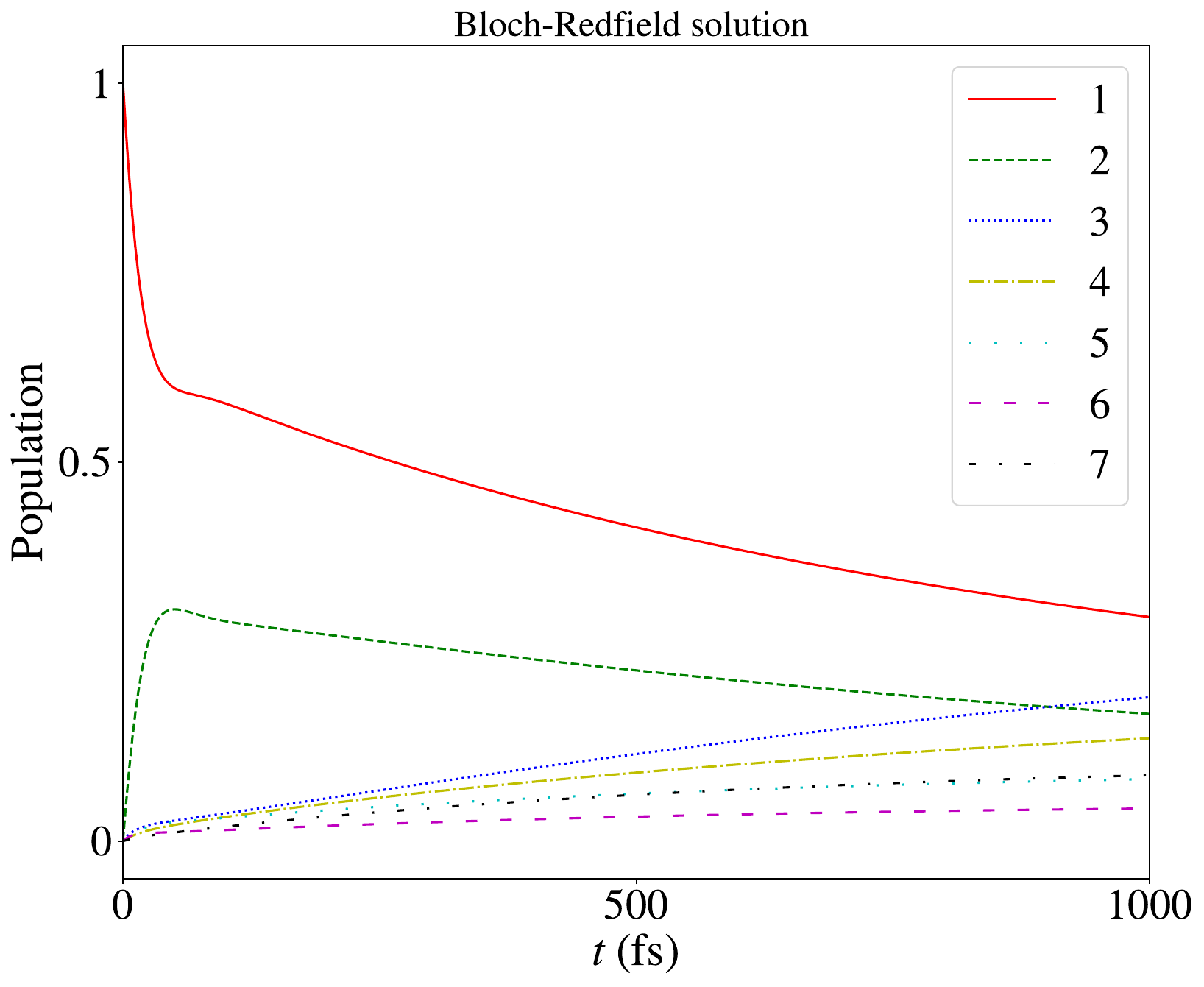}
\caption{Site-populations for the FMO complex versus time as given by the standard Bloch-Redfield solver in QuTiP. Parameters are the same as in \figref{fig4}: $\lambda = 35~\mathrm{cm}^{-1}$, $\gamma^{-1} = 166$ fs, $T=300$ K.}\label{fig5}
\end{figure}

To show how a non-Markovian environment differs from a Markovian one, in \figref{fig5} we show the results, using the same parameters, for the standard Bloch-Redfield solver in QuTiP. This solver, under standard Born-Markov-Secular approximations, takes into account the frequency dependence of the bath spectral density, the eigenstate structure of the system Hamiltonian, and the pure dephasing that arises from the low-frequency behavior of the bath. 

In the accompanying notebook 2 \cite{examindex} we explicitly show how these different contributions affect the dynamics, and we see there quite explicitly that the pure dephasing (in the eigenbasis) has a very strong effect on the dynamics, largely suppressing coherent oscillations. This comparison explicitly shows that weak-coupling models fail to predict the correct coherence timescale in this case, due to the highly non-Markovian environment \cite{Akihito09,melina} (as determined partially by the narrow cut-off $\gamma$).  In earlier works suppression of low-frequency noise was shown to arise for super-Ohmic spectral densities \cite{Kreisbeck2012}, based on arguments regarding the frequency dependence of the Markovian pure-dephasing contribution, but  appears here due to the non-Markovian nature of the environment.

\begin{figure}[]
\includegraphics[width = \columnwidth]{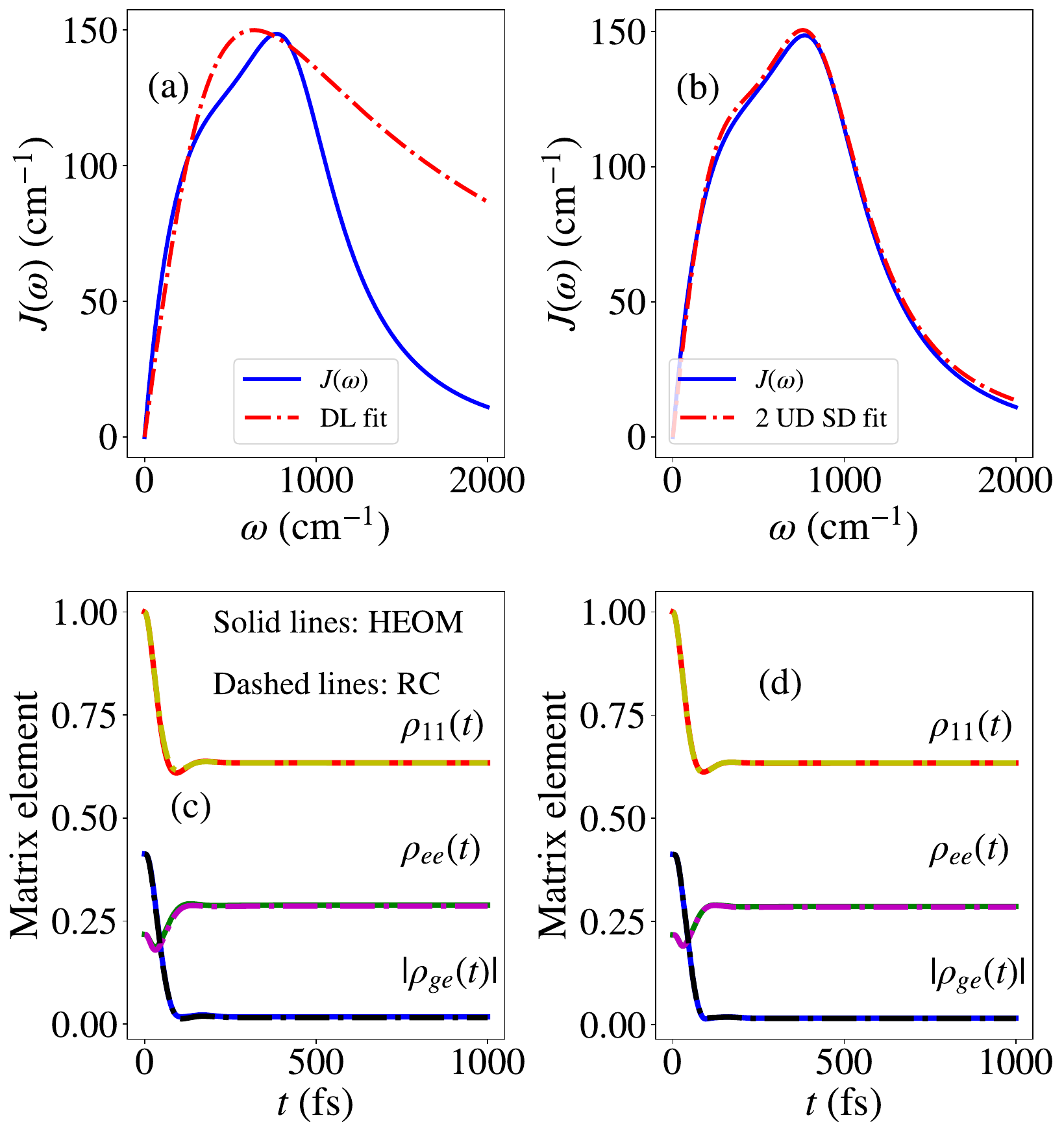}
\includegraphics[width = \columnwidth]{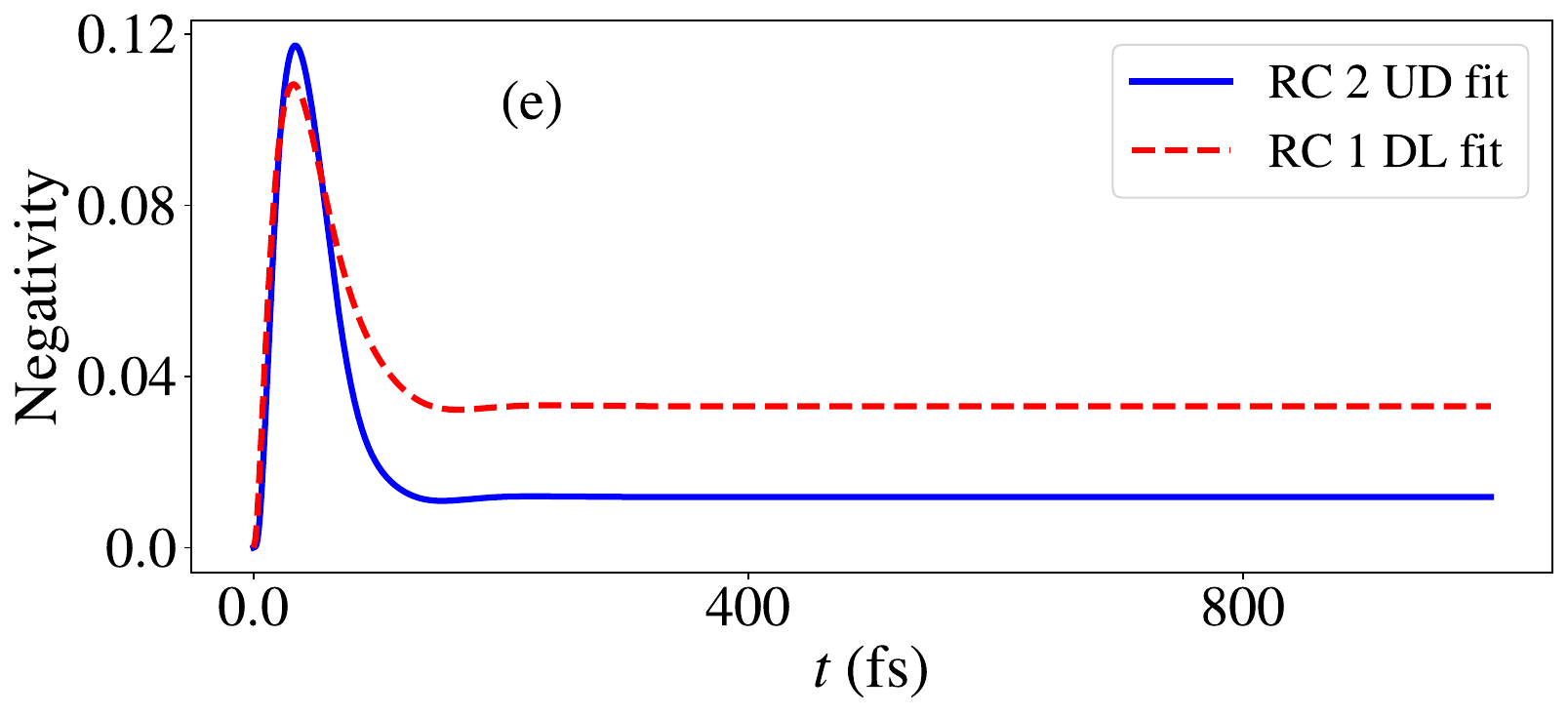}
\caption{{\color{red} (a) shows a fit of the spectral density $J(\omega)$ from \eqref{pnasj} with a single over-damped Drude-Lorentz spectral density function, giving $\gamma = 637$ cm$^{-1}$ and $\lambda = 300 $ cm$^{-1}$. (b) shows a fit instead using two under-damped brownian motion spectral densities, giving $\alpha_1^2 = 0.11\omega_{c,1}^3$, $\omega_{c,1} = 900$ cm$^{-1}$, $\Gamma_1 = 800$ cm$^{-1}$, and $\alpha_2^2 = 0.25\omega_{c,2}^3$, $\omega_{c,2} = 500$ cm$^{-1}$, $\Gamma_2 = 990$ cm$^{-1}$. (c) shows the HEOM and RC results obtained for dynamics of several matrix elements on the two-site dimer model using the fit from (a), while (d) shows the same using the fit from (b).  The RC and HEOM produce comparable results, and the choice of fitting function has little influence on the system dynamics in this case. (e) shows the entanglement between system and RC mode(s) as determined by the negativity using the single RC mode model for the fit in (a) and and two RC modes using the fit in (b). }} \label{figOD}
\end{figure}

{\color{red} Interestingly, whether the electronic coherence, as exhibited in the oscillations seen in \figref{fig4},  actually survives in nature is still hotly debated \cite{Duan8493}.  In \cite{Duan8493} the authors observe no electronic coherence beyond $60$ fs, in contrast with earlier experiments \cite{Engel2007,Panitchayangkoon2010,chengreview,Collini2010}. In modelling their data they employ a theoretical method using a much broader and stronger bath spectral density than that used in \cite{Akihito09} that strongly suppresses the coherent oscillations of the type shown in in \figref{fig4}. Here we adapt the spectral density employed in the supplementary information of \cite{Duan8493}:
\beq
J(\omega) = \gamma_{\mathrm{Ohm}} \omega e^{-\omega/\omega_c^{\mathrm{Ohm}}} + \frac{2}{\pi}\frac{S\Omega_{\mathrm{r}}^3\omega\Gamma_{\mathrm{r}}}{(\Omega_{\mathrm{r}}^2-\omega^2)^2 + \omega^2 \Gamma_{\mathrm{r}}^2}\label{pnasj}
\eeq
 with the parameters  listed there \footnote{For the Ohmic part $\gamma_{\mathrm{Ohm}}=0.7$, $\omega_c^{\mathrm{Ohm}} = 350$ cm$^{-1}$, and for the resonant part $S=0.12$, $\Omega_{\mathrm{r}}=900$ cm$^{-1}$ and $\Gamma_{\mathrm{r}} = 700$ cm$^{-1}$} (see \fig{figOD}(a) and (b) for a plot of this function).
 
To gain some insight into this parameter regime, we first truncate the FMO space to only the first two-sites coupled to a single bath. This gives a reduced total Hamiltonian similar to our earlier examples,
\beq
H_{\mathrm{red}}= \frac{\epsilon}{2} \sigma_z + \Delta \sigma_x
\eeq
with a coupling operator to a single bath $Q=\sigma_z/\sqrt{2}$, where the factor $\sqrt{2}$ arises in the reduction of the original two-bath model to the normal modes of a single-bath model (see \cite{Iles16} for more details on this step in the reduction).  Using the FMO parameters we employed earlier we find $\epsilon = -120$ cm$^{-1}$ and $\Delta  = -87.7$ cm$^{-1}$, and again use $T=300$ K.

We then simulate this reduced system using the HEOM in two ways: first we do a very rough approximation using a single overdamped Drude-Lorentz spectral density, then we do a more nuanced fit using two underdamped spectral densities.   We then repeat the simulations using the reaction-coordinate (RC) approach \cite{iles2014environmental}, a method that is more approximate than the HEOM but known to perform well at high temperatures \cite{lewis} even with broad spectral densities.

Importantly, the RC method gives us access to a density operator for the original system plus an effective collective ``reaction coordinate'' mode, from which we can obtain system-environment properties less directly accessible with the HEOM. In addition, since the RC approach amounts to performing a Bogoliubov transformation on the original bath modes combined with tracing out a residual environment, we can calculate the system-RC entanglement via quantities like the negativity. We expect this to be a lower-bound on the entanglement between the system and the original full environment (since said entanglement cannot be increased by local operations and classical communication), if the perturbative approximation for deriving a master equation for the residual bath modes holds.

Figure~\ref{figOD} demonstrates the results from these various approaches. First of all, we use the exact HEOM method to validate the more approximate RC method in this parameter regime, and we see that for both choices of spectral densities the methods produce identical system population dynamics and steady states, and that both spectral density choices produce essentially the same result. This appears to be because the large coupling at low frequencies dominates the system dynamics and causes quick suppression of coherent dynamics in both the site [$\rho_{11}$ in \figref{figOD}(c) and (d)] and eigenbases [$\rho_{ge}$ in \figref{figOD}(c) and (d)].

Furthermore, \fig{figOD}(e) shows the result of calculating the negativity between the system and the single RC mode used for the overdamped Drude-Lorentz spectral density, and the two modes used for the two-underdamped spectral densities. We observe dynamics in this quantity up to several $100$~fs and significant non-zero steadystate values in both cases.  Note that the non-secular master equation used to describe the residual bath of the reaction-coordinates can sometimes induce non-positive eigenvalues in the RC modes themselves, as it does not necessarily preserve positivity. This potentially signals the start of a breakdown of the perturbative approximation needed to derive this master equation, and thus suggests a limit on how much we can infer about the original system-bath entanglement. However, for the parameters used in \figref{figOD}, negative eigenvalues only appear at very short times and are small compared to the negativity values shown in \fig{figOD}(e).

In a more complex model involving more sites, we expect the negativity between the first two sites and the environment to decay with time as the electronic excitation moves through the complex. Nevertheless these results suggest that observing electronic coherence alone can be insufficient to determine whether non-trivial system-environment correlations can be neglected, particularly when the effective coupling to the environment is on the order of the system energies.  In addition, we see here the benefit of combining the exact HEOM with more approximate but nuanced methods, like the RC approach.  
}

\subsection{Example 3: Quantum heat transport}\label{sec:therm}

In the examples discussed so far, we have investigated the time evolution of the system density matrix alone.
An intriguing feature of the HEOM is that it also allows us to study observables that depend on the state of the environment, which is encoded in the auxiliary density operators (ADOs).
In this example, we will demonstrate this feature by calculating the quantum heat current in a two two-level system setup.
The heat transport in this setup was originally studied by Kato and Tanimura in Ref.~\cite{KatoJChemPhys2015}; we here show that their results can be reproduced with ease using our framework.

The setup consists of two coupled two-level systems, which are each in turn coupled to an individual heat reservoir.
The Hamiltonian describing the two-level systems is
\beq
	H_S = \frac \epsilon 2 \bigl( \sigma_z^1 + \sigma_z^2 \bigr) + J_{12} \bigl( \sigma_+^1 \sigma_-^2 + \sigma_-^1 \sigma_+^2 \bigr) ,
\eeq
where $\epsilon$ denotes the level splitting, $J_{12}$ the coupling between two-level systems, and $\sigma^K_{z,\pm}$ denotes the usual Pauli matrices for the $K$-th two-level system.
Each reservoir is modeled as a continuum of harmonic modes with an overdamped Drude-Lorentz spectral density as defined in \eqref{JD}.
The multi-reservoir setup can be easily treated using the HEOM by including a bath index $K \in \{ 1, 2 \}$ in the multi-index of the ADOs.
The bath coupling operators are given by $Q_K = \sigma_x^K$.

A temperature difference $T_1 > T_2$ induces heat transport in the setup.
Due to the second law of thermodynamics, the heat flow is on average directed from the hot side to the cold one.
Our goal is to determine the individual heat currents between the system and the reservoirs, which are generally given by $j_K(t) = \partial_t \langle H_B^K \rangle$ with $H_B^K$ being the Hamiltonian of the $K$-th reservoir.
These can be determined from the ADOs as follows \cite{KatoJChemPhys2016}:
\beq
	j_K &=& \Gamma_T^K \operatorname{tr}\bigl[ [[H_S, Q_K], Q_K]\, \rho \bigr] - 2 C_I^K(0) \operatorname{tr}\bigl[ Q_K^2 \rho \bigr] \nonumber \\
	&+& \sum_{j = R,I} \sum_{k = 1}^{N_j^K} \gamma_{k}^{j;K} \operatorname{tr}[ Q_K \rho^{(jk;K)} ] . \label{eq:heat_current}
\eeq
We here added indices signifying the reservoir-dependency to the rates $\gamma_k^{j;K}$, the cut-offs $N_j^K$, the correlation function $C^K(t)$ and the Tanimura-terminator $\Gamma_T^K$.
The symbol $\rho^{(jk;K)}$ denotes the level-$1$ ADO corresponding to the multi-index $(0 \dots 0, 1, 0 \dots 0)$ with only a single non-zero entry at the indicated position.

\begin{figure}
    \includegraphics[width=\columnwidth]{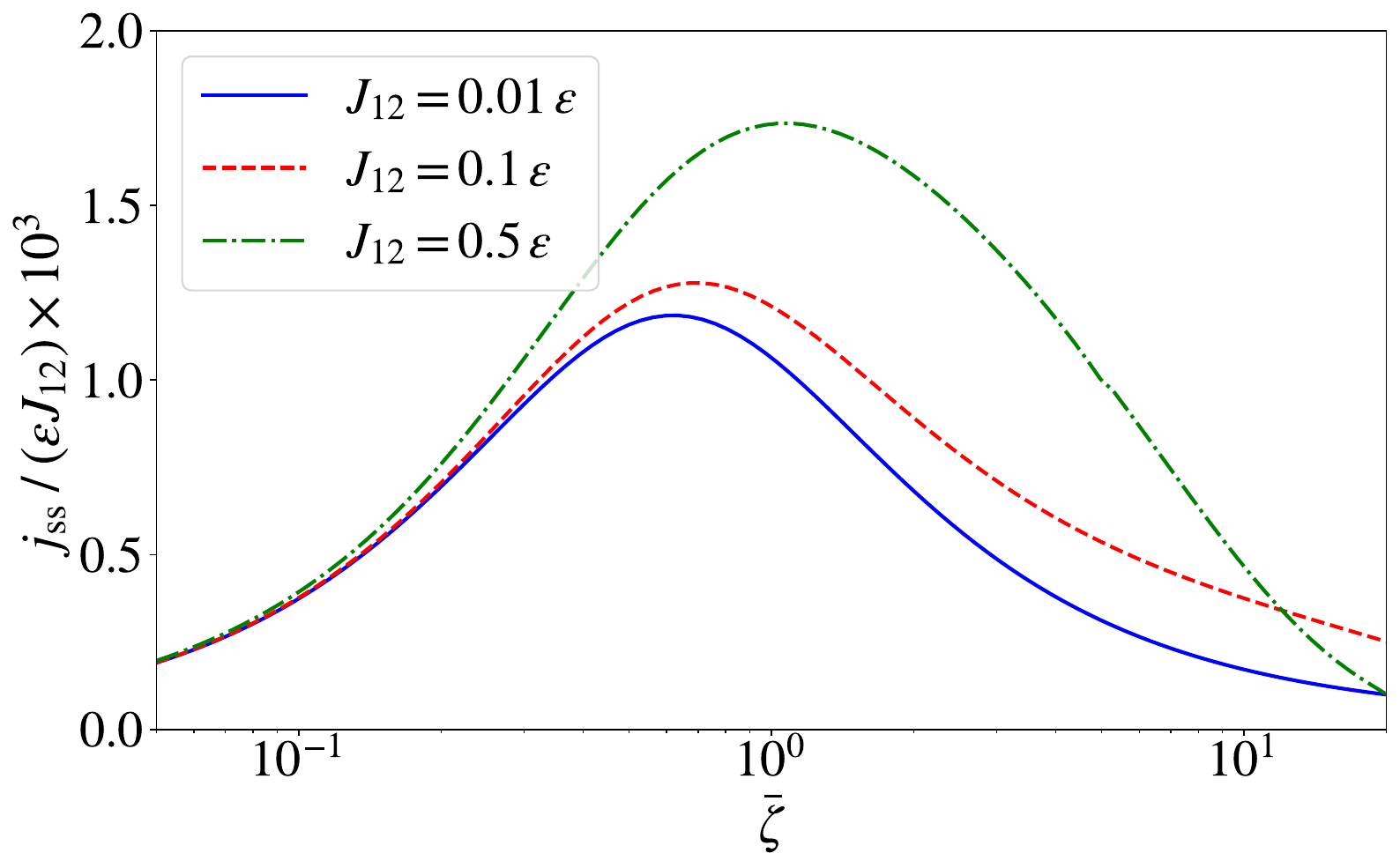}
    \caption{
        HEOM results for the steady-state heat current in the two two-level system setup.
        The parameters were chosen as follows: $\gamma_1 = \gamma_2 = 2\epsilon$, $T_1 = 2.02\epsilon$, $T_2 = 1.98\epsilon$ and $\lambda_1 = \lambda_2 = \bar\zeta (J_{12} / 2)$, where the dimensionless parameter $\bar\zeta$ parametrizes the system-bath coupling strength.
        This plot is a reproduction of the HEOM-curves in Figs.~3(a-i) to 3(a-iii) of Ref.~\cite{KatoJChemPhys2015}. The code for generating this figure can be found in example notebook 3 \cite{examindex}}.
    \label{fig:heat}
\end{figure}

In our library, the \code{HierarchyADOsState} class facilitates the extraction of specific ADOs from the full state of the simulation.
The formula in \eqref{eq:heat_current} can therefore be easily implemented, as we demonstrate in detail in example notebook 3 \cite{examindex}.
In Fig.~\ref{fig:heat}, we show the behavior of the steady-state heat current $j_{\vphantom{2}\text{ss}} = j_{2,\text{ss}} = -j_{1,\text{ss}}$ in the high-temperature regime as a function of the system-reservoir coupling strength.
The figure shows that the heat current vanishes both for large and small coupling strengths and reaches a maximum in the intermediate regime.
The behavior at large coupling strengths is caused by a quantum Zeno-like effect, where a strong system-reservoir interaction suppresses the coherence of the system, inhibiting the energy flow between the individual two-level systems \cite{KatoJChemPhys2015,taka1,taka2}.

\subsection{Example 4: Driven systems and dynamical decoupling}\label{sec:dd}

Another powerful aspect of the HEOM lies in the fact that it is valid for any system Hamiltonian, even time-dependent ones. As a concrete example, we show here how it can be used to study dynamical-decoupling, a common tool used to ``undo'' dephasing from an environment, even for finite pulse duration (i.e., away from the "bang-bang" pulse limit).

The general idea of dynamical decoupling is that dephasing from a non-Markovian environment can be undone by the application of certain choices of fast pulse-control on the system itself. We follow the protocol outlined in [\onlinecite{ViolaDD}], which involves rapid applications of $\pi$ pulses, interspersed with periods of interaction with the environment.  The intuitive concept is that if these pulses act faster than the memory time of the bath, by swapping the state of the two-level system, the effect of the environment can be reversed due to a change in sign in the interaction. 

\begin{figure}[ht]
\includegraphics[width = \columnwidth]{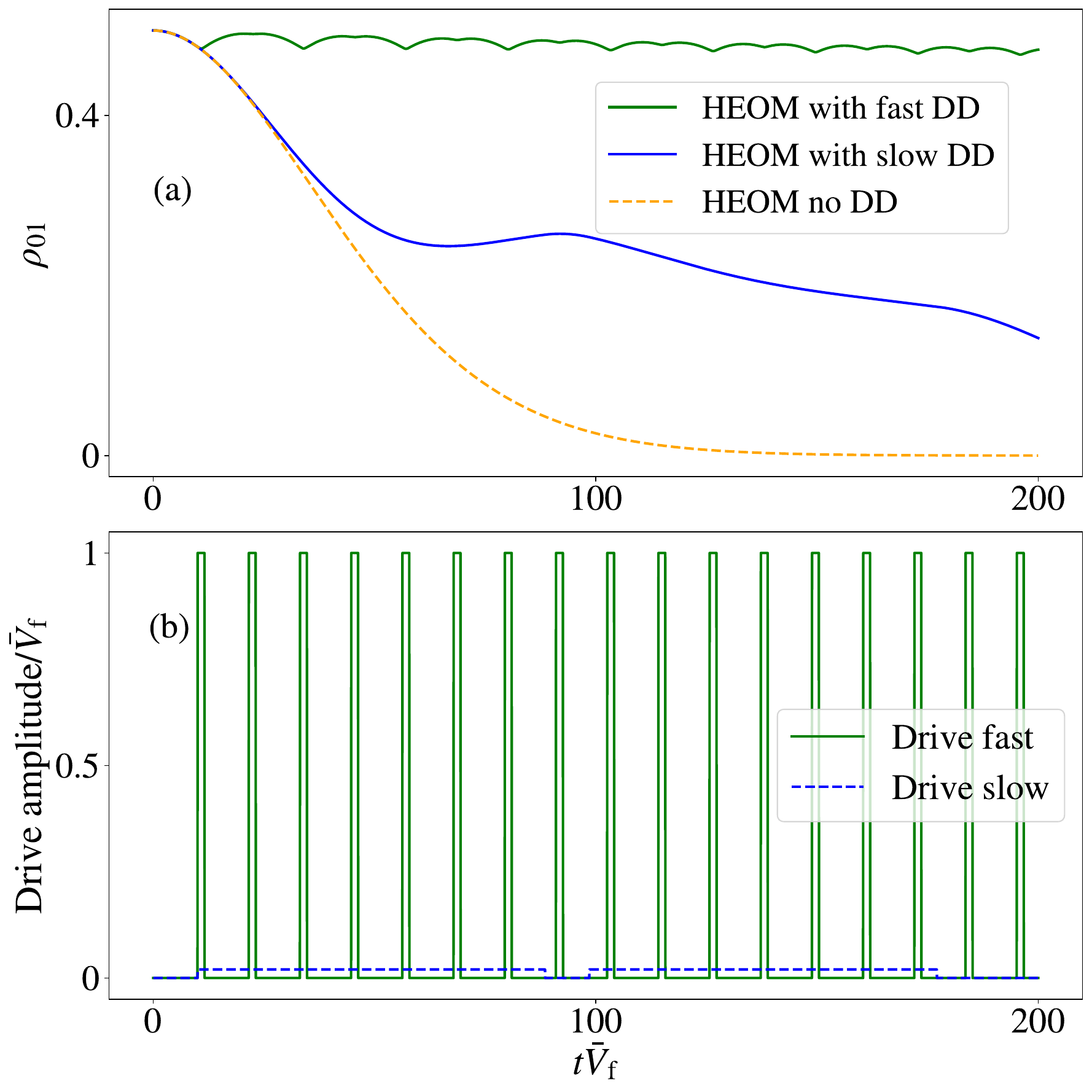}
\caption{Example of dynamical decoupling showing the coherence $\rho_{01}$ of a two-level system coupled to an environment with a Drude-Lorentz spectral density. Here $\Delta = 0 $, $\lambda = 1\times 10^{-4} \bar{V}_{\mathrm{f}}$, $\gamma= 1\times 10^{-2} \bar{V}_{\mathrm{f}}$, $T=0.1\bar{V}_{\mathrm{f}}$, where $\bar{V}_{\mathrm{f}}$ is the amplitude of the fast drive. For the slow drive we choose $\bar{V}=0.02 \bar{V}_{\mathrm{f}}$, and in both cases we choose a pause period of $\Delta_t\bar{V}_{\mathrm{f}} = 10 $. The lower figure, (b), shows the drive amplitudes as a function of time.  In (a) we see that very slow pulses perform much worse than fast pulses, as expected due to the competition with the bath memory time.  Note that here and in \figref{udd} we operate in a rotating frame with the drive, and assume that the system frequency is much larger than the drive amplitude ($\epsilon \gg \bar{V}$) such that the rotating wave approximation is valid. The code for generating this figure can be found in  example notebook 4 \cite{examindex}.}\label{DDfig}
\end{figure}

Here we choose again a Hamiltonian in the form of \eqref{HS}, but setting $\Delta = 0$ and adding a time-dependent drive term,
\beq
H_{\mathrm{S}} &=& \frac{\epsilon}{2}\sigma_z + H_D(t).
\label{HS2}
\eeq
We use the same coupling operator as \eqref{Q}, and the same spectral density as \eqref{JD}.  We move to an interaction picture  with respect to the time-independent part of $H_S$, such that the drive Hamiltonian has the form,
\beq
\tilde{H}_D(t) = \sum_{n=1}^{n_p} V_{n}(t)\, \sigma_x \, .
\eeq
 The pulse is chosen to be, for equal spacing $\Delta_t$ between pulses, $V_{n}(t)= \bar{V}$, for $n\Delta_t + (n-1) \tau_p \leq t \leq n \Delta_t + n\tau_p$, and zero elsewhere,  (see \figref{DDfig}b for a schematic of the pulse shapes). The period of the actual pulses themselves is $\tau_p \bar{V} =\pi/2$.  
Note that in comparison to [\onlinecite{ViolaDD}] we omit some phase factors that appear if one considers a more realistic model of the drive. 

In \figref{DDfig}a (see also example notebook 4 \cite{examindex}) we show the time-evolution of the coherence of the two-level system assuming the initial condition $\psi(t=0)=\frac{1}{\sqrt{2}}(\ket{0} + \ket{1})$.  When the $\tau_p$ is chosen to be fast (and  the corresponding $\bar{V}$ strong so that $\bar{V}\tau_p = \pi/2$) we see the expected cancellation of dephasing occurring (green curve).  If the pulse is too slow, we see the onset of dephasing (blue curve).  The decoherence expected with no pulse at all is also shown (orange dashed curve).

It has been shown \cite{gotz07prl,renbaoPRL} that for environments with very sharp cut-offs \cite{Uhrig_2008} (e.g., step-functions), a squared-sinusoidal choice of spacing is optimal (a scheme referred to as Uhrig Dynamical Decoupling (UDD)). In that case, one sets $V_{n}(t)= \bar{V}$ when 
\beq&&
\sin^2[\pi n/(2n_p+2)](T_{\mathrm{max}}-n_p\tau_p) + (n-1) \tau_p \leq t \nonumber \\
&\leq& \sin^2[\pi n/(2n_p+2)](T_{\mathrm{max}}-n_p\tau_p) + n\tau_p,\eeq and zero elsewhere.  This produces pulses that are clustered more closely together at the beginning and end of the overall time-evolution period $T_{\mathrm{max}}$.  

\begin{figure}[ht]
\includegraphics[width = \columnwidth]{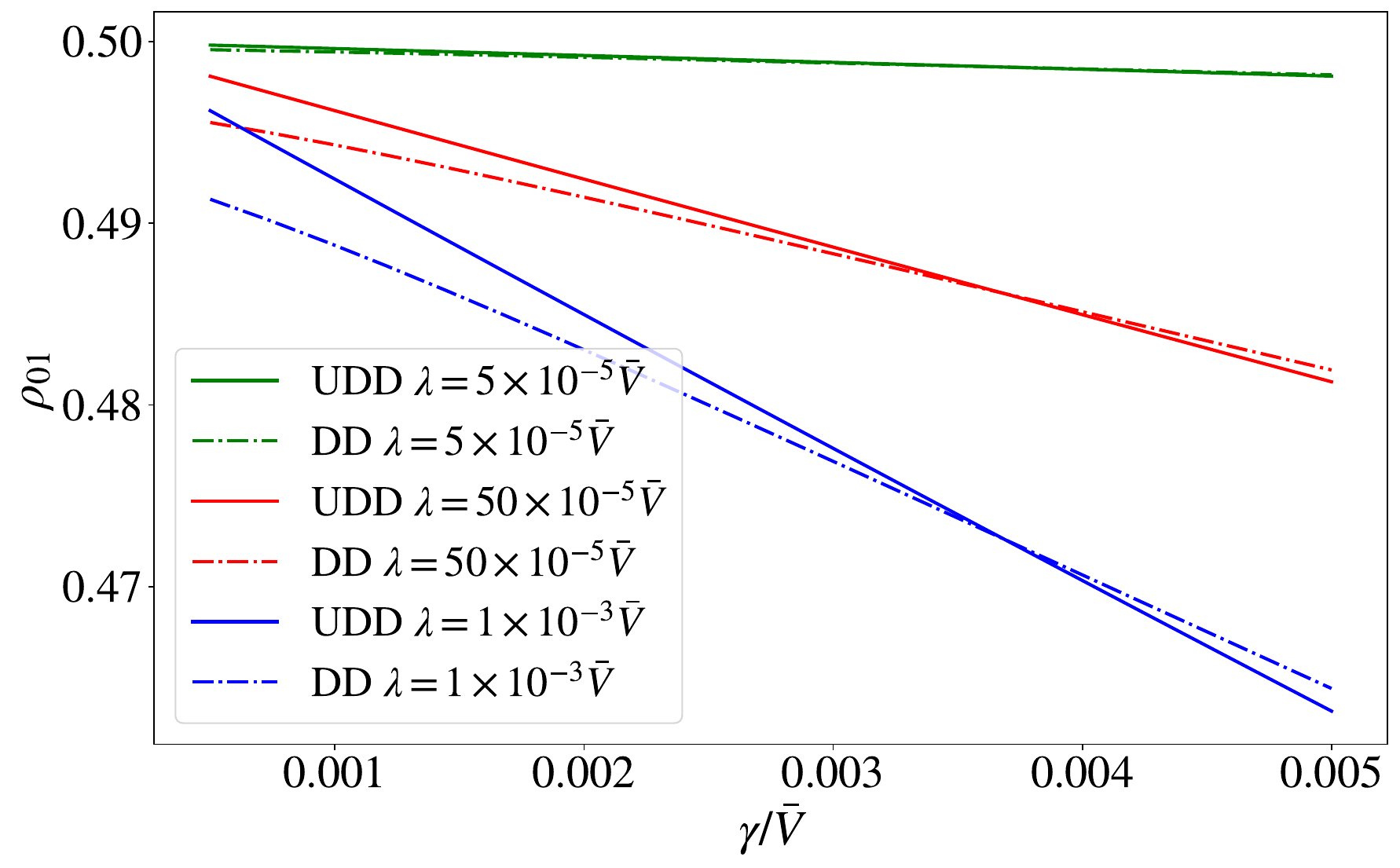}
\caption{Example of dynamical decoupling showing the coherence $\rho_{01}$, after a total evolution time $T_{\mathrm{max}}=1000 (1/\bar{V})$, of a two-level system coupled to an environment with a Drude-Lorentz spectral density versus $\gamma$, the width (inverse memory time) of the bath. Each run includes 100 dynamical decoupling pulses, and we compare equally spaced pulses (labelled DD in the figure legend) with the UDD scheme described in the text. The parameters here are chosen as  $\Delta = 0 $, $\lambda = 0.001\bar{V}, 0.0005\bar{V}, 0.00005\bar{V}$, and $T=0.05\bar{V}$. There is a clear crossover for larger values of $\gamma$ where the UDD scheme becomes less efficient than the standard equally spaced scheme (DD). The code for generating this figure can be found in example notebook 4 \cite{examindex}. }\label{udd}
\end{figure}

The Drude-Lorentz spectral density we use for this example has a very long tail, and thus this approach is not guaranteed to be optimal at all times.  In \figref{udd} we demonstrate this by comparing the final coherence $\rho_{01}$ at time $t=T_{\mathrm{max}}$ as a function of $\gamma$ and for $100$ decoupling pulses. We see that for smaller $\gamma$ the UDD outperforms the equally spaced case, but as we increase $\gamma$ its performance drops.  Interestingly, as expected from the discussions in \cite{gotz07prl, renbaoPRL}, the coupling strength $\lambda$ does not reduce the relative effectiveness of the UDD appreciably. These results demonstrate how HEOM can be used to validate control schemes in realistic scenarios such as finite pulse length and realistic choices of spectral density.

To evaluate time-dependence in the Hamiltonian, our code mirrors the typical approach taken for standard solvers used in the QuTiP package.  The time-dependence is defined through a function that is sent alongside the dependent part of the Hamiltonian in a list.  For the above example, this follows simply as
\begin{center}
\code{H = [Hsys, [sigmax(), drive]]}
\end{center}
where \code{drive} is a function determining the time-dependence. 
The various parameters can also be defined as arguments for the function.  Our code also accepts \code{QobjEvo} format for the time dependence.

\subsection{Arbitrary spectral densities}\label{sec:fitting}

A great deal of effort has been made in the literature to optimize the analytical decomposition of the Drude-Lorentz spectral density into exponentials, so as to minimize the numerical overhead in the HEOM.  For more general spectral densities, one can consider two approaches: fitting the spectral density directly with Drude-Lorentz or underdamped spectral densities, or evaluating the correlation functions and fitting those directly.  Both approaches have been discussed in the literature \cite{fittingeisfeld, shifit}, with the general consensus being that the optimal choice depends on what one is evaluating -- the dynamics of short or long time scales, the spectral response of the system, etc.  A more recent method also suggests generalizing the types of  functions used in the decomposition of the correlation functions \cite{ikeda2020}.

To compare the different fitting approaches, we can consider a class of spectral densities with polynomial frequency dependence and exponential cut-off:
\beq
J(\omega) = \alpha \omega^s\omega_c^{1-s}\exp\left(-\frac{\omega}{\omega_c}\right)
\label{Jo}
\eeq

The correlation functions for these spectral densities are \cite{BrandesDiss}

\beq
C(t) &=& \frac{1}{\pi} \alpha \omega^{1-s}\beta^{-(s+1)}\Gamma(s+1)\\
&&\left[ \zeta\left(s+1,\frac{1+\beta\omega_c - i\omega_c t}{\beta\omega_c}\right)+\right.\nonumber\\
&&\left. \zeta\left(s+1,\frac{1+ i\omega_c t}{\beta\omega_c}\right)\right]\nonumber\label{co} \, ,
\eeq
wherein $\Gamma$ is the Gamma function and $\zeta$ is the generalized  Zeta function.

\begin{figure}[]
\includegraphics[width = \columnwidth]{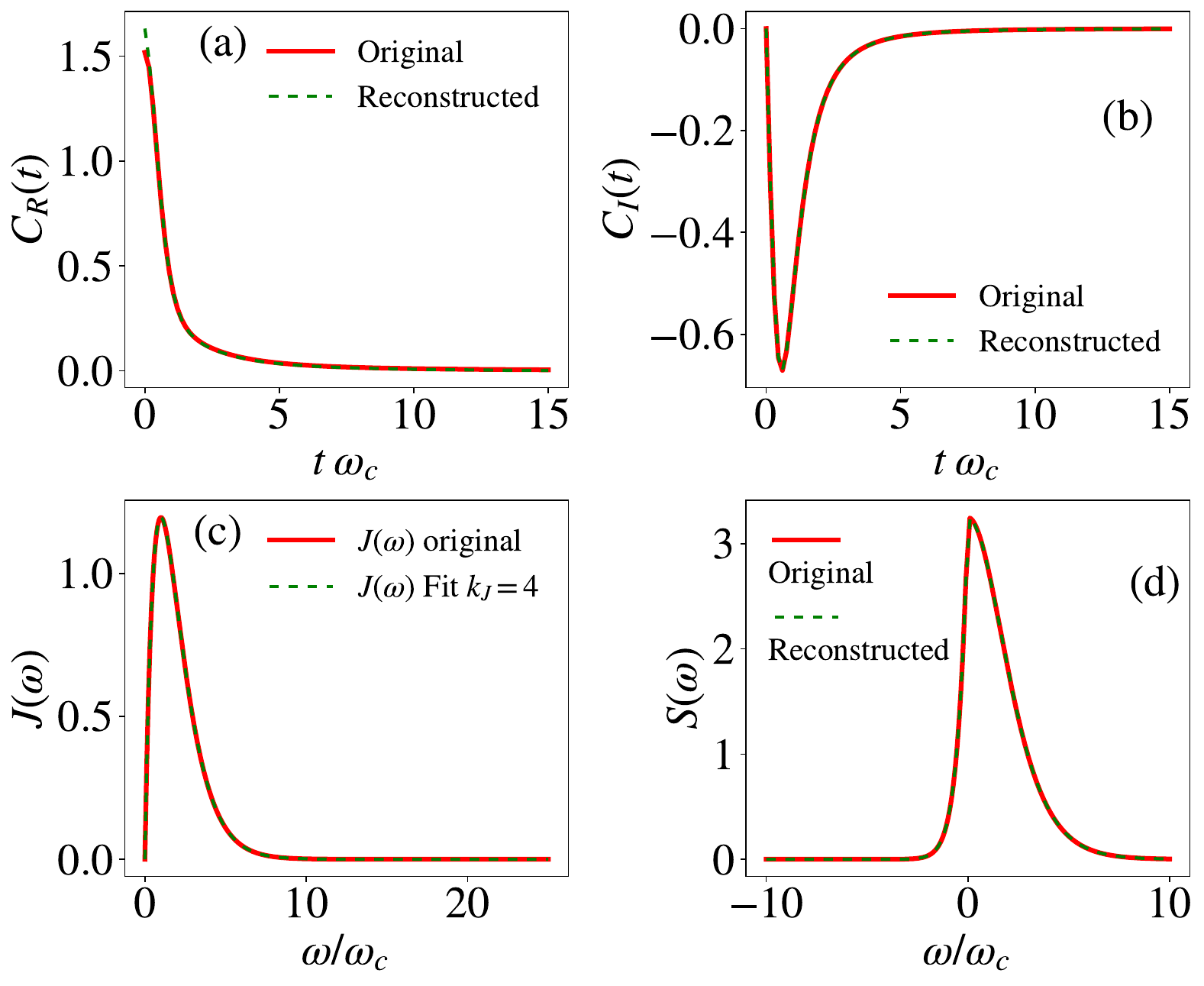}
\caption{A demonstration of the fitting of the spectral density of an Ohmic environment with exponential cut-off [\eqref{Jo} with $s=1$] using $k_J =4$ underdamped Brownian motion terms. We use bath parameters   $\alpha = 3.25$ and $T=\omega_c$. The resulting correlation functions are shown in (a) and (b), the original and fit spectral density is shown in (c), and the resulting power spectrum is shown in (d). In evaluating the correlation functions we use only one Matsubara term $K=1$. The residual error seen in the real part is then compensated for using a terminator. The code for generating this figure can be found in example notebook 1d \cite{examindex}.}\label{fig8}
\end{figure}

Here we present one simple example of fitting the Ohmic case $s=1$, but the following can also be applied to other examples.
If one chooses to fit \eqref{Jo} directly, Meier and Tannor \cite{tannor-meier} proposed using a set of  underdamped spectral densities in the form
\beq
J_U^F(\omega) =  \sum_{i=1}^{k_J}\frac{2\alpha_i^2 \Gamma'_i \omega}{\left[\{(\omega + \Omega_i)^2 + {\Gamma'}_i^2\}\{(\omega - \Omega_i)^2 + {\Gamma'}_i^2\}\right]} \, .
\eeq
Here, in converting this decomposition into the same form as those in Section \ref{sec:spin_boson_model}, we set $\Gamma_i'= \Gamma_i/2$.  In \figref{fig8} (see also  example notebook 1d \cite{examindex}) we show an example of such a  fitting, for an Ohmic environment $s=1$, using $k_J =4$ terms. With the fitting parameters we can also reconstruct the correlation functions and power spectrum, as a check of validity of the fit.   

Instead, if we choose to fit the correlation functions, we treat the real and imaginary terms separately, and expand them as 
\beq
C_R^F(t) &=& \sum_{i=1}^{k_R} c_R^ie^{-\gamma_R^i t}\cos(\omega_R^i t)\nonumber\\
C_I^F(t) &=& \sum_{i=1}^{k_I} c_I^ie^{-\gamma_I^i t}\sin(\omega_I^i t)
\label{fit}
\eeq

\begin{figure}[]
\includegraphics[width = \columnwidth]{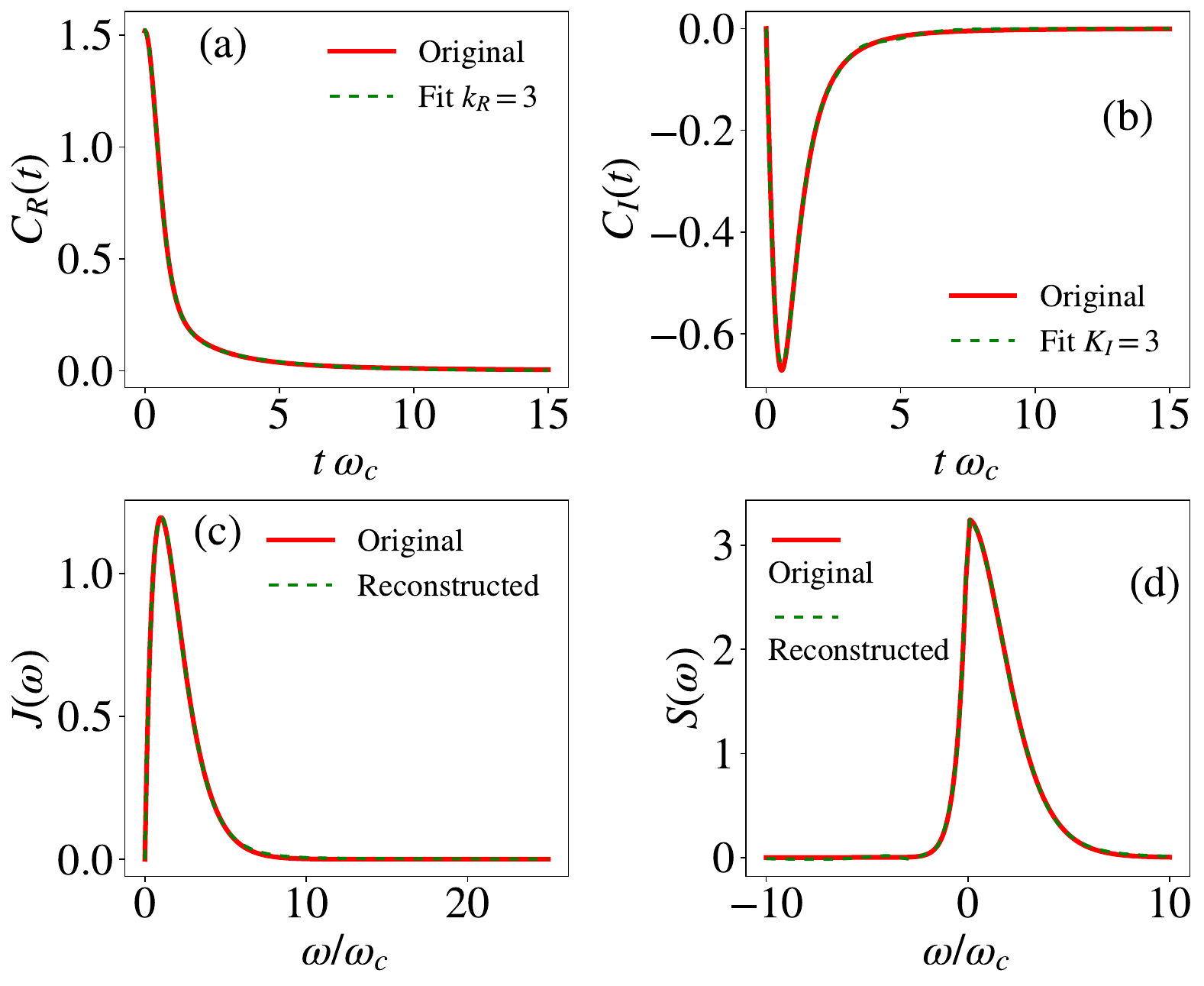}
\caption{A demonstration of the fitting of the correlation functions of an Ohmic environment with exponential cut-off [\eqref{Jo} with $s=1$] using $k_R =3$ and $k_I=3$ exponential terms terms. We use bath parameters $\alpha = 3.25$ and $T=\omega_c$.  The original and fit correlation functions are show in (a) and (b). The resulting spectral density and power spectrum are also shown in (c) and (d) respectively. The code for generating this figure can be found in example notebook 1d \cite{examindex}.}\label{fig9}
\end{figure}

In \figref{fig9} we show the fitting for the same parameters as \figref{fig8} using $k_R=k_I =3$ terms, as well as the reconstructed power spectrum. One can see that at negative frequencies this fit has some residual oscillations, which may lead to some small error in the dynamics.  In both spectrum and correlation function fitting, it is important to restrain the fitting parameters to some degree, as arbitrary parameters can make achieving convergence in the resulting HEOM simulation difficult.

\begin{figure}[]
\includegraphics[width = \columnwidth]{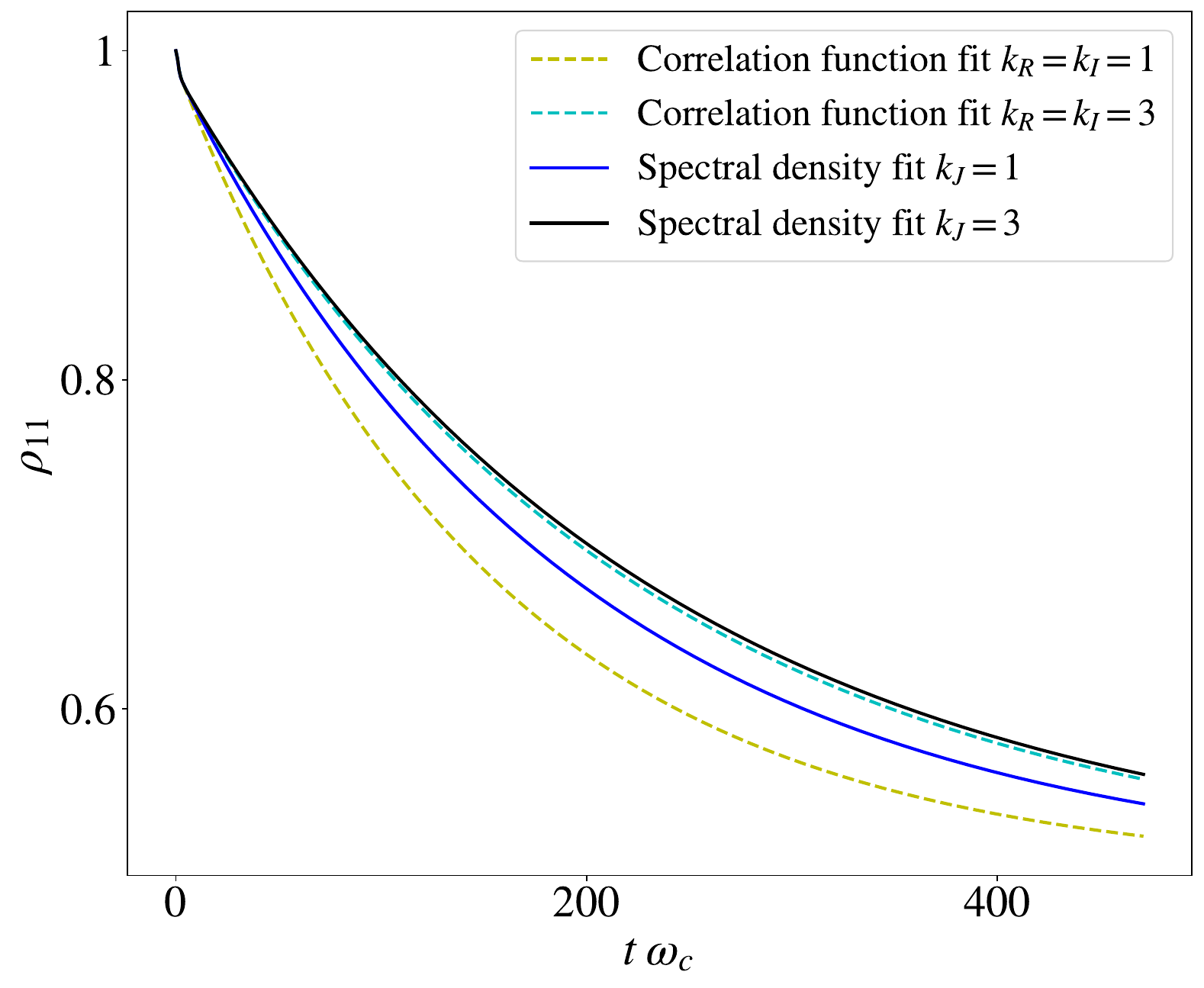}
\caption{Dynamics of the excited state
population $\rho_{11}(t)$ for a two-level system coupled to an Ohmic environment with exponential cut-off [\eqref{Jo} with $s=1$] for $\alpha = 3.25$, $T=0.5\omega_c$, $\epsilon=0$, $\Delta = 0.2 \omega_c$.  We show results for fitting the spectral density with $k_J =1$, and $3$ underdamped Brownian motion terms, and for fitting the correlation functions with $k_r=k_I=1$ and $3$ terms.  For the case where we fit the spectral density, we use $N_k=1$ and the standard terminator to approximate the other Matsubara terms. Here, and in \figref{fig7b} a HEOM truncation of $N_c=11$ is used to get convergence. Examples of  these fits used to generate the data in this figure are shown in \figref{fig8} and \figref{fig9}. The code for generating this figure can be found in example notebook 1d \cite{examindex}.}\label{fig7}
\end{figure}

\begin{figure}[]
\includegraphics[width = \columnwidth]{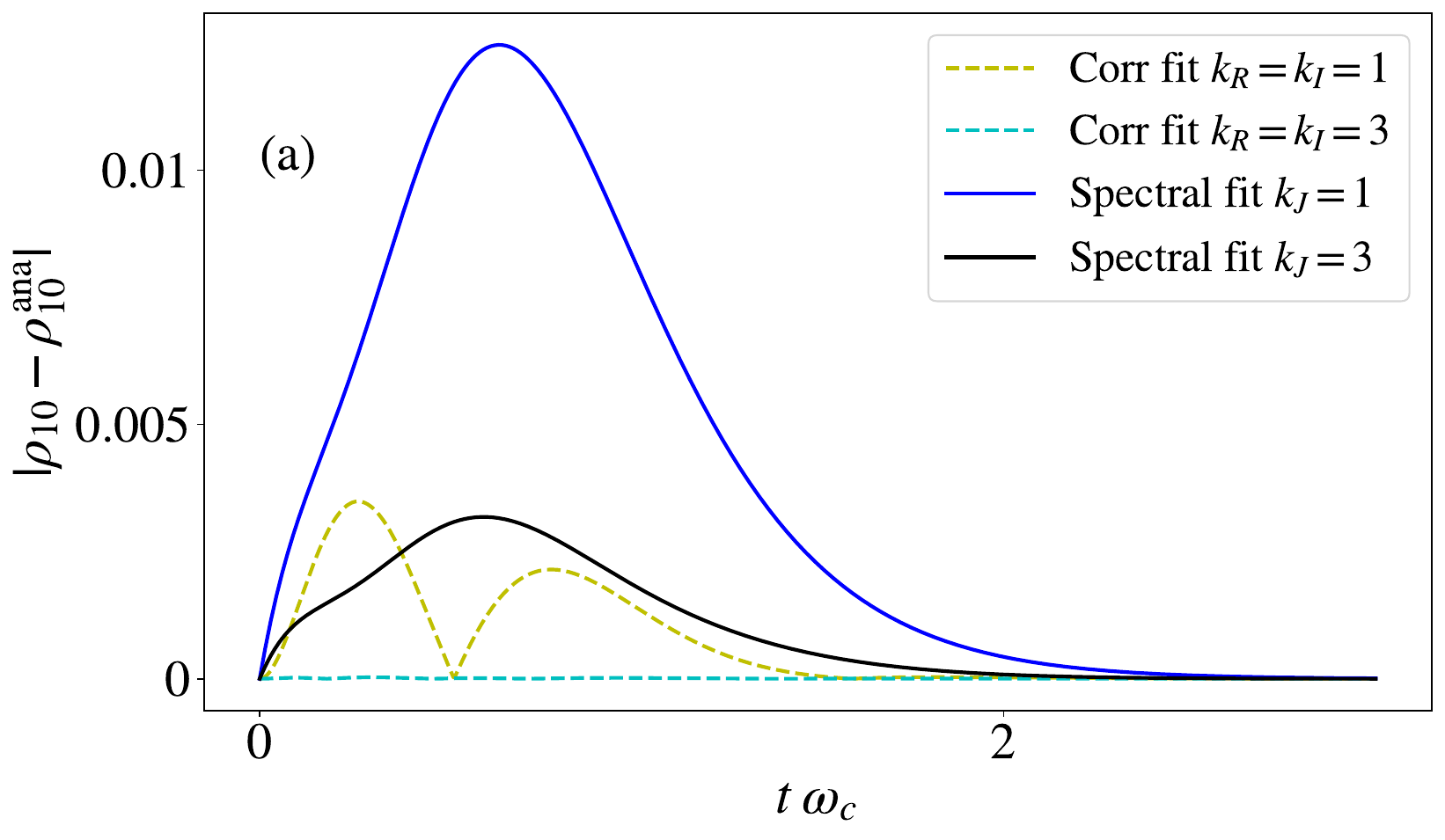}
\includegraphics[width =
\columnwidth]{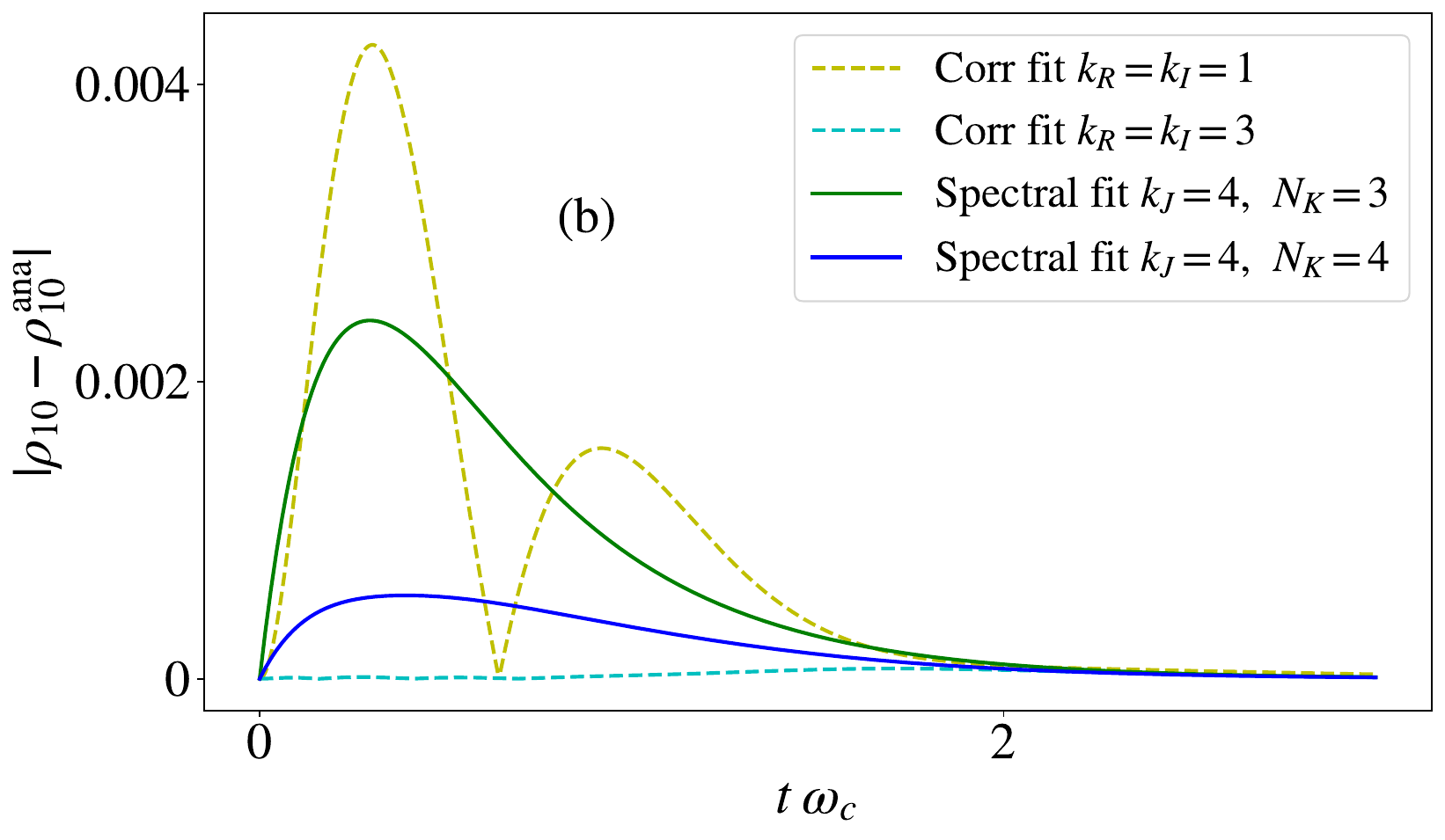}
\caption{{\color{red} Difference in the dynamics as evaluated by the HEOM and that given by an analytical result for the coherence $|\rho_{10}(t)-\rho_{10}^{\mathrm{ana}}|$ for a two-level system initially prepared with maximal coherence, and coupled to an Ohmic environment with exponential cut-off [\eqref{Jo} with $s=1$] for $\alpha = 3.25$, $\epsilon=0$, and $\Delta = 0$. The latter means that, unlike \figref{fig7}, this result is integrable, and we can compare to an analytical result. (a) is for $T=0.5\omega_c$, as in \figref{fig7}, while (b) is for a lower temperature of $T=0.25\omega_c$. In (a) we show results for fitting the spectral density with $k_J =1$, and $3$ underdamped Brownian motion terms, and for fitting the correlation functions with $k_r=k_I=1$ and $3$ terms.  For the case where we fit the spectral density, we here use $N_k=1$ and the standard terminator to approximate the other Matsubara terms. In (b) the lower temperature means we have to both increase the number of spectral functions and employ more Matsubara terms to reach results comparable to the correlation function fitting results.}}\label{fig7b}
\end{figure}

We compare the results of both approaches in \figref{fig7} for parameters close to \figref{fig2}, except with slightly narrower bath and longer time-scales.  More importantly, for the spectral density decomposition, we employ a terminator to compensate for the relatively low truncation of the Matsubara terms (see \figref{fig8}).  {\color{red} The number of ADOs, and hence numerical cost, with the correlation fitting method is $2(k_R+k_I)$ . For the spectral density fitting each $k_J$ term adds two exponents for the non-Matsubara term, and  a single exponent for each matsubara term. Hence the number of exponents is $k_J(2 + N_K)$.   In \figref{fig7} we see that $k_R=k_I=3$ gives comparable results to $k_J=3, N_k=1$ ($N_k=0$ performs badly and is not shown), implying similar numerical cost for approximately the same accuracy in this case (with slightly less exponents needed for the spectral fitting approach). 

However, as one lowers the temperature, more and more Matsubara terms are needed in the spectrum approach. This is corroborated by \figref{fig7b} which compares the fitting parameters in the HEOM against the analytical result available when we set the system Hamiltonian to commute with the bath coupling operator (here achieved by setting $\Delta=0$). First we see that, in \figref{fig7b}(a), for the same parameters and temperature as \figref{fig7}, for $k_J=3$ and $k_R=k_I=1$ we see similar magnitude of accuracy, while $k_R=k_I=3$ does exceptionally well. In \figref{fig7b}(b), at a temperature lowered by a factor of a half compared to \figref{fig7}, we find that the spectral decomposition requires both more spectral terms and more Matsubara terms to be on par with the correlation function fitting result. In particular, the cyan curve for $k_R=k_I=3$ requires twelve exponents in total, while the blue curve for $K_J=4$ and $N_K=4$ requires 24 exponents, suggesting much lower numerical cost for correlation function fitting in the low temperature regime.

In all cases we use a standard least-squares algorithm for the fitting procedure. More sophisticated fitting approaches \cite{shifit}, or taking advantage of known properties of the bath correlation functions, may produce better overall results. One may also consider fitting the power spectrum as a third method (which equates to fitting the Fourier transform of the correlation functions). Also, while one can use analytical limits in some case to benchmark the results (like the pure dephasing result in \figref{fig7b}), in general one may lack a way validate any given result. One potential strategy is to perform both fitting procedures (spectrum and correlation functions), and check convergence of the system dynamics in both cases (large differences between them suggest one or both contain non-negligible errors). A more formal refinement of this approach may be to develop error bars on a given result based on the different fitting procedures used, akin to the error bars used in [\onlinecite{Lambert2019}] based on noise introduced into the fitting procedure.
}

\section{Fermionic Environments}\label{sec:ferm}

\subsection{Basic Definitions}

In the previous section we summarized how a bath of bosonic modes can influence a discrete system. Another common scenario is that of a macroscopic conductor coupled to a microscopic impurity, in which case one must consider how discrete states interact with a continuum of fermionic modes.  Such a scenario is traditionally analyzed with a range of many-body techniques, but in recent years it was shown that the HEOM method can be applied to this scenario as well.  Here we summarize the basic definitions, and present some examples from the fields of quantum transport and single-molecule electronics.  

Again, in the standard second-quantized Hamiltonian formalism, we can write the interaction of a fermionic mode $c$, $\{c,c^{\dagger}\} = 1$, with support on the system space, with a fermonic bath,
\beq
H = H_{\mathrm{S}} + \sum_k \varepsilon_k d_k^{\dagger}d_k +  \sum_k f_k \left(d_k^{\dagger} c + c^{\dagger} d_k\right) \;,
\label{Hferm}
\eeq
where, as with the bosonic case, $H_S$ is the free Hamiltonian of the system (which can contain arbitrary degrees of freedom) and $H_{\mathrm{B}} = \sum_k \varepsilon_k d_k^{\dagger}d_k$ is the free Hamiltonian of the electronic reservoir. 
 
As with the bosonic case, we can characterize the free environment through its second-order correlation functions. However, now we discriminate between the order in which excitations are created or destroyed in the environment, so that we define two different correlation functions, defined by the choice of $\sigma=\pm$:
\beq
C^{\sigma}(t) = \frac{1}{2\pi} \int_{-\infty}^{\infty} d\omega e^{\sigma i \omega t} J(\omega) f_F[\sigma\beta(\omega - \mu)] \; .
\label{Cf}
\eeq
In the following we primarily follow the notation and choices made by Schinabeck {\em et al.} \cite{schinabeck2}.  Unlike in the bosonic case, we do not divide the correlation functions explicitly into real and imaginary parts, but in the form
\beq
C^{\sigma}(t) \approx \sum_{l=0}^{l_{\mathrm{max}}} \eta_{\sigma,l} e^{-\gamma_{\sigma,l}t}
\eeq
where $\eta$ and $\gamma$ can be complex numbers. 

The hierarchical equations of motion for such an environment \cite{Yan08,Li_2012,schinabeck2} are then written as
\beq
\dot{\rho}^{(n)}_{j_n...j_1} &=& \left(\mathcal{L} - \sum_{m=1}^n \gamma_{j_m}\right)\rho^{(n)}_{j_n...j_1} 
- i \sum_j\mathcal{A}^{\bar{\sigma}}{\rho}^{(n+1)}_{jj_n...j_1} \nonumber\\
&-& i \sum_{m=1}^n(-)^{n-m}\mathcal{C}_{j_m}\rho^{(n-1)}_{j_n...j_{m+1}j_{m-1}...j_1} \; ,
\label{Hf}
\eeq
where now $j = (\sigma,l)$ is a multi-index, $\mathcal{L}=-iH_S^{\times}$ is the normal system evolution, and as before $\rho^{(0)}$ is the density operator of the system and $n>0$ are auxiliary densityoperators (ADOs) .  The superoperators are
\beq
\mathcal{A}^{\bar{\sigma}}\rho^{(n)} &=& c^{\bar{\sigma}}\rho^{(n)} + (-)^n\rho^{(n)}c^{\bar{\sigma}}\\
\mathcal{C}^{\sigma}\rho^{(n)} &=& \eta_{\sigma,l}c^{\sigma}\rho^{(n)} - (-)^n\rho^{(n)}\eta_{\bar{\sigma},l}^*c^{{\sigma}} \; .
\eeq

Here we use $\bar{\sigma}=-\sigma$, $c^{+}=c^{\dagger}$, and $c^{-} = c$.  The above definitions are easily extended to multiple baths by extension of the index $j$, as in the bosonic case. Still following the notation of [\onlinecite{schinabeck2}] in this generalization, we will refer to the multi-index $j=(K,\sigma,l)$, such that there are $N_j=2N_K(l_{\mathrm{max}}+1)$ indices where $N_K$ are the number of reservoirs, and the factor of two arises for the two signs of $\sigma$, and $(l_{\mathrm{max}}+1)$ are the number of exponents in the decomposition of the reservoir correlation functions.

The primary differences between the bosonic and fermionic HEOM are the following.  Firstly, in the fermionic HEOM multi-indices $j$ can only take the values $0$ or $1$, due to the fermionic nature of the operators in the environment, and thus the truncation parameter $N_C$ is the maximum amount of non-zero indices in $j$, and $N_C\leq N_j$ ( equality implies no truncation).
Secondly, because fermionic operators anti-commute, the ordering of the indices is now important. Hence the various parity terms that arise in \eqref{Hf} depending on how many non-zero indices are in the list.  This is, in particular, vital for the second term in \eqref{Hf}, where on raising an index value from $0$ to $1$, one can notice that we append the new non-zero index to the left of the list, ${\rho}^{(n+1)}_{ {\bf  j}j_n...j_1}$. Thus one must then move the index to its appropriate position in the unique multi-index $j$. In doing so, we pick up an additional parity on this term depending on how many non-zero terms appear between the position on the left and default position of said ADO in the list. Note that in other implementations \cite{HEOMQUICK}, an additional optimization is done by taking advantage of the hermiticity relation between ADOs with conjugate $\sigma$.  We do not implement that optimization at this stage.

To demonstrate the application of this method, we employ a Lorentzian spectral density of the form
\beq
J(\omega) = \frac{\eta  W^2 }{[(\omega-\mu)^2 +W^2 ]} \; ,
\label{Jf}
\eeq
where $\mu$ is the chemical potential, $W$ is the width of the environment, as before, and $\eta$ a coupling strength.
For fermions, the distribution in \eqref{Cf} is
\beq
f_F (x) = [\exp(x) + 1]^{-1} .
\label{Ff}
\eeq
We again have the choice of decomposing the correlation functions into exponentials in several different ways. The Matsubara decomposition proceeds similarly to the bosonic case. Thus, here we present the Pad\'e decomposition, which approximates the Fermi distribution as 
\beq
f_F(x) \approx f_F^{\mathrm{approx}}(x) = \frac{1}{2} - \sum_l^{l_{\mathrm{max}}} \frac{2k_l x}{x^2 + \epsilon_l^2} \; ,
\eeq
where $k_l$ and $\epsilon_l$ have to be evaluated numerically, and depend on the choice of $l_{\mathrm{max}}$ (see [\onlinecite{BetterPade}],  and the example notebooks 5a and 5b \cite{examindex}, for details).  Performing this decomposition, and evaluating the integral for the correlation functions gives (see supp. info. of \cite{schinabeck2})
\beq
\eta_{0} = \frac{\eta W}{2} f_F^{\mathrm{approx}}(i\beta W) \; ,
\eeq
\beq
\gamma_{\sigma,0} = W - \sigma i\mu \; ,
\eeq
\beq
\eta_{l\neq 0} = -i\cdot \frac{k_m}{\beta} \cdot \frac{\eta W^2}{-\frac{\epsilon^2_m}{\beta^2} + W^2} \; ,
\eeq
\beq
\gamma_{\sigma,l\neq 0}= \frac{\epsilon_m}{\beta} - \sigma i \mu \; .
\eeq 
The Pad\'e decomposition we used for the bosonic case follows in a similar way.


\subsection{Observables}

As with the bosonic method, one can directly solve for the dynamics and steady-state properties of system observables.  Generally in transport problems, for which the fermionic method is particularly useful, other quantities of interest are the steady-state current, conductance, higher-order transport statistics, and spectral properties. The current, for example, depends on properties of the environment, and can be extracted from the auxiliary density operators, rather than just the system density matrix.  It is related to the first-order ADOs (see section 2.2.4 of [\onlinecite{phdthesis}] for a detailed derivation):

\beq
\ex{I_K} = -i e \sum_{l} \mathrm{Tr_S}\left[c \rho^{(1)}_{K,+,l} - c^{\dagger}\rho^{(1)}_{K,-,l}\right] \; .
\eeq

As with the heat-engine examples discussed earlier, the HierarchyADOsState class allows the extraction of specific ADOs from the results. These can then be straightforwardly used with the above formula (see example notebooks 5a and 5b \cite{examindex}).

\subsection{Code Functionality}

The functionality of the fermionic solver is largely the same as the bosonic one. Each bath is specified via its decomposition into correlation functions, and the associated system operator coupling operator should be provided as a single operator associated with the coupling to the $d_k$ modes in the bath (i.e., the single fermion $c$ in the examples below). We provide a Matsubara (\code{LorentzianBath}) and Pade decomposition (\code{LorentzianPadeBath}) of the Lorentzian spectral density given in \eqref{Jf}, or a generic bath (\code{FermionicBath}).  Examples of this functionality can be found in the documentation and the example notebooks \cite{examindex}. 

\subsection{Example 1: Integrable single-impurity model} \label{sec:resonant}

To benchmark the accuracy of the code it is useful to consider an integrable example of the single-impurity Anderson (SIAM) model.  In this case we consider a single spin-less fermionic impurity coupled to two reservoirs
\beq
H_{\mathrm{SIAM}} &=& c^{\dagger}c + \sum_{K=L/R,k} \varepsilon_{K,k} d_{K,k}^{\dagger}d_{K,k} \nonumber\\
&+&  \sum_{K=L/R,k} f_{K,k} \left(d_{K,k}^{\dagger} c + c^{\dagger} d_{K,k}\right).
\label{Hsiam}
\eeq
We assume the reservoirs are described by \eqref{Jf} and \eqref{Ff} but generalize them to have different chemical potentials so that the impurity sees a bias $\Delta\mu = \mu_L - \mu_R$.  The steady-state current is a well-known analytical result \cite{brandes},
\beq
\!\!\ex{I} = \int_{-\infty}^{\infty}\!\!\frac{2d\omega }{\pi}\frac{J_L(\omega)J_R(\omega)[f\{\beta(\omega-\mu_L)\}\!-\!f\{\beta(\omega-\mu_R)\}]}{[J_L(\omega)+J_R(\omega)]^2+4[\omega-\epsilon \!- \lambda_L(\omega)\!-\lambda_R(\omega)]^2}\nonumber
\eeq
where the Lamb shifts are 
\beq
\lambda_{L}(\omega) = \frac{(\omega-\mu_{L})J_{L}(\omega,\mu_{L})}{2W}.\\
\lambda_{R}(\omega) = \frac{(\omega-\mu_{R})J_{R}(\omega,\mu_{R})}{2W}.
\eeq

\begin{figure}[]
\includegraphics[width = \columnwidth]{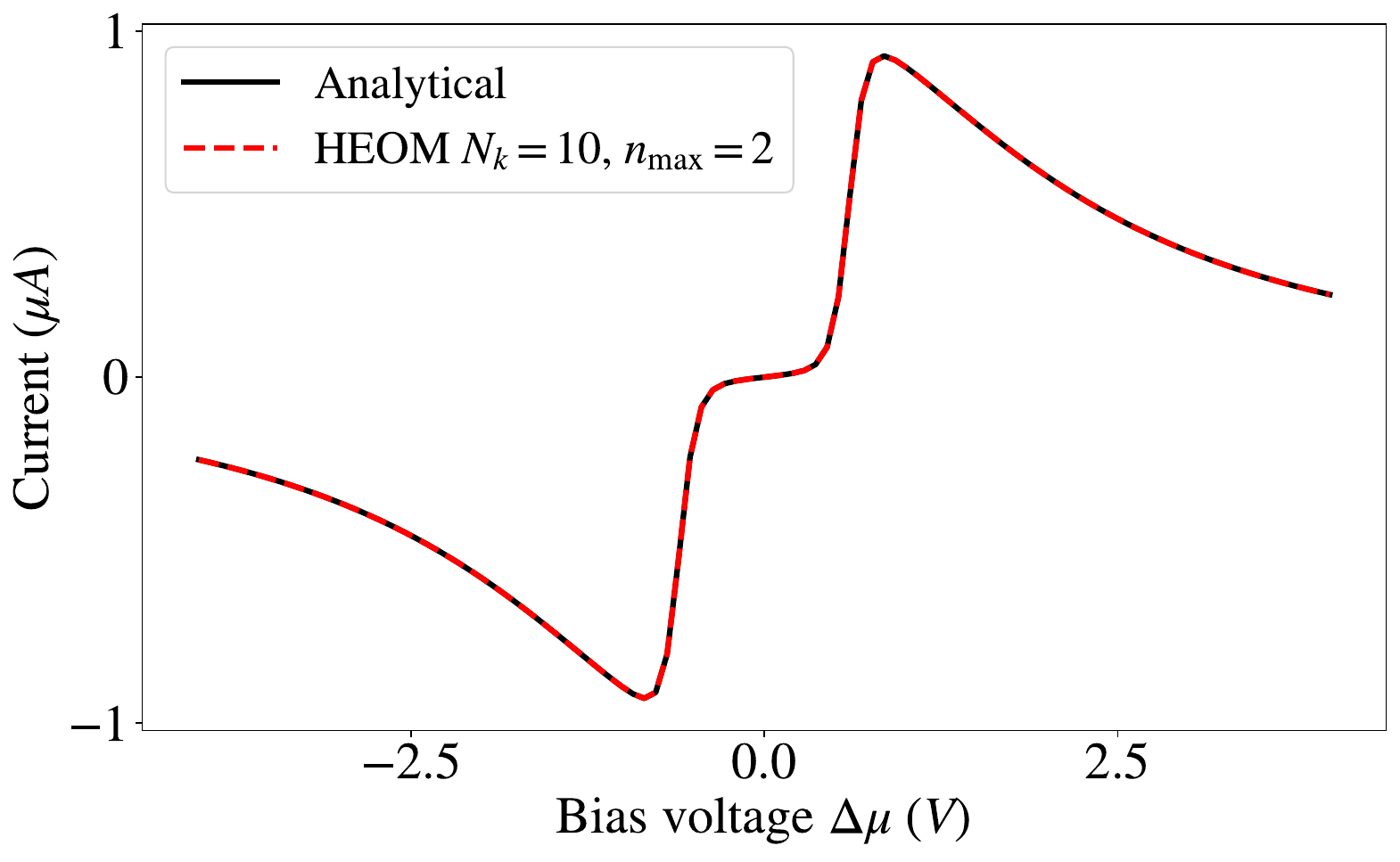}
\caption{ The analytical current and the result from the HEOM, using the Pad\'e decomposition, for the single-impurity model, as a function of the bias voltage $\Delta \mu$, for $\eta = 0.01$ eV, $T=300$ K, and $W=1 $ eV, and the impurity energy is $\epsilon = 0.3$ eV.  The code for generating this figure can be found in example notebook 5a \cite{examindex}. }\label{fig10}
\end{figure}

In \figref{fig10} we show the analytical current and the result from the HEOM, using the Pad\'e decomposition, as a function of the bias voltage $\Delta\mu$, for $\eta = 0.01$ eV, $T=300$ K, and $W=1 $ eV, with the impurity energy $\epsilon = 0.3$ eV.  From this figure we see quite clearly that the current flow through the impurity is zero until the bias window is equal or larger than $\epsilon$.  The current then follows quite closely the energy dependence of the reservoir power spectrum multiplied by the Fermi function.  We note that, as discussed in-depth in [\onlinecite{fermtrunc}], the HEOM here converges at $n_{\mathrm{max}} = 2$.

\subsection{Example 2: Vibronically assisted transport}\label{sec:resonanceandboson}

Moving away from this simple, integrable, case, we consider the example of a single impurity coupled to a vibronic mode. Such a model is widely discussed in the literature of single-molecule electronics and nanosystems \cite{schinabeck,Jakub2020}.  Here we explicitly reproduce an example from figure 1(a) in [\onlinecite{schinabeck2}].  The model used there considers \eqref{Hsiam} with the addition of an explicit vibronic degree of freedom in terms of a single bosonic mode:

\beq 
H_{\mathrm{vib}} = H_{\mathrm{SIAM}} + \Omega a^{\dagger}a + \lambda (a+a^{\dagger})c{^\dagger}c.
\eeq
In solving this example one has several options: treat the bosonic mode as a HEOM bosonic environment (as done in [\onlinecite{schinabeck}]), include it explicitly in the system Hamiltonian with a discrete Fock space representation, perform a polaron transformation first (as done in [\onlinecite{schinabeck2}]), or finally, perform a polaron transform and absorb the vibronic coupling into the lead couplings (which is then treated with an additional approximation, as done in [\onlinecite{jiang}] and [\onlinecite{Jakub2020}]).  We choose the second option here, and truncate the bosonic Fock space at a level $N$ which gives convergence. It is equivalent to the first and third options, while being the most general approach for the demonstration of our library, and is also numerically more difficult than the polaron and direct HEOM approach. 

\begin{figure}[]
\includegraphics[width = \columnwidth]{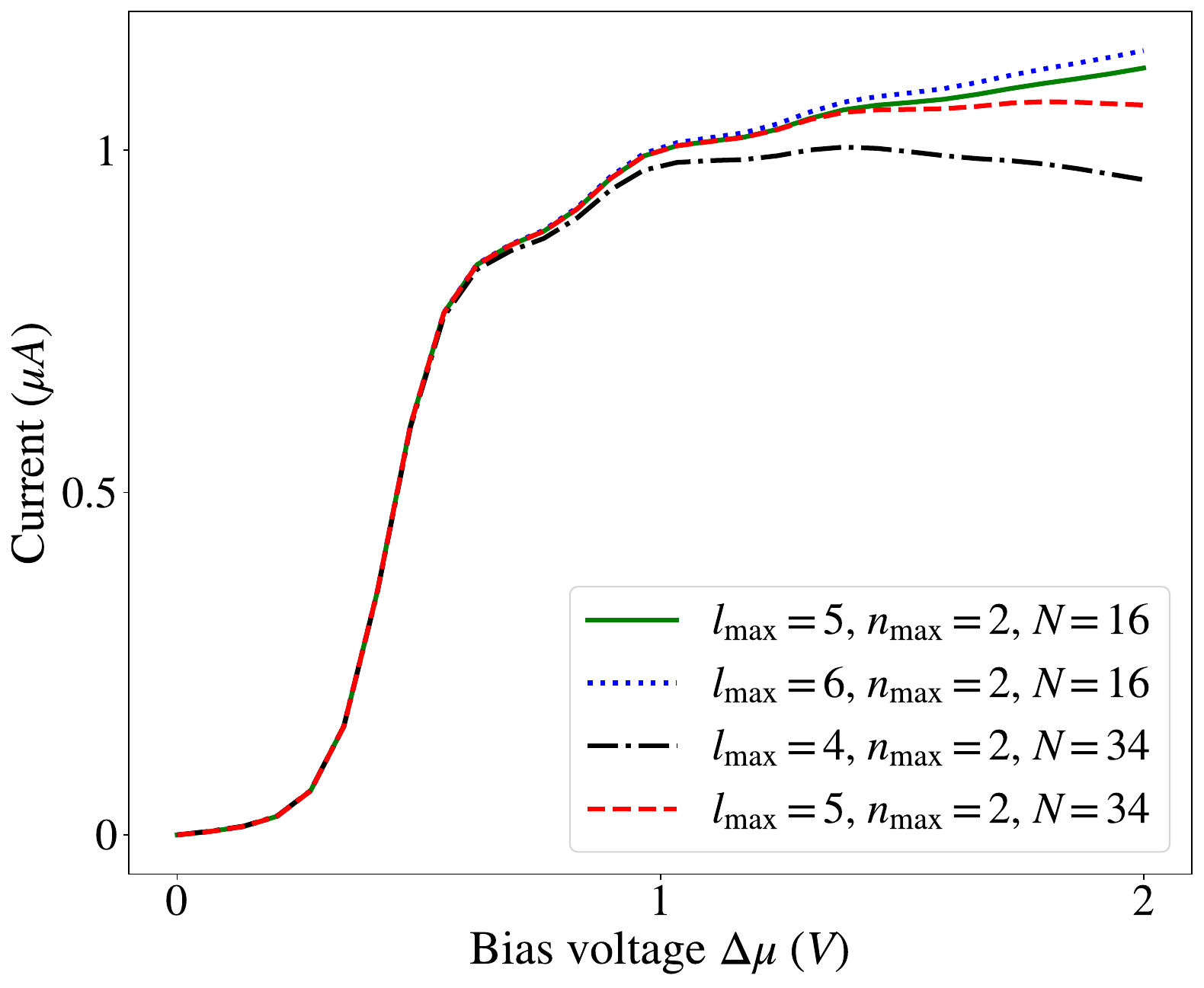}
\caption{Current versus the bias voltage for a single impurity coupled to a vibronic mode with $\epsilon=0.3$, eV $\Omega = 0.2$ eV, $\lambda = 0.12$ eV, $T=300$ K, $W=10^4$ eV, and $\eta = 0.01$ eV. Note that increasing $l_{\mathrm{max}}$ tends to increase the current at large bias voltages, while increasing the other convergence parameter, the bosonic cut-off $N$, tends to decrease it, gradually giving a converged result in the vicinity of the red curve. The code for generating this figure can be found in example notebook 5b \cite{examindex}.}\label{fig11}
\end{figure}

In \figref{fig11} we show an example of the current versus the bias voltage for $\epsilon=0.3$, eV, $\Omega = 0.2$ eV, $\lambda = 0.12$ eV, $T=300$ K, $W=10^4$ eV, and $\eta = 0.01$ eV.  We see that at large bias voltages we need a large number of bosonic Fock states to see convergence in the current, consistent with the appendix of [\onlinecite{schinabeck2}].  The result appears to converge to the result reported in that work for $l_{\mathrm{max}}=5$ and $N=34$ and $N_C=2$.  Note that, if one instead employs the polaron transform approach, a higher-tier result (e.g., $N_C=3$ here) can be obtained at little numerical cost because its affect on the tier below can be found analytically. We do not employ this optimization  here, however, as it is not generally applicable to all problem Hamiltonians.  Interestingly, while the current seems to converge, the correlation functions obtained with the Pad\'e decomposition  are not converged (due to the wide-band limit of $W=10^4$ eV). This ``quicker'' convergence of system properties occurs because, for fermionic environments, certain highly over-damped contributions to the bath correlation functions do not contribute to the system dynamics (see the supplementary information of \cite{cirio2022,Cirio2021} for a detailed proof).

\section{Conclusions}
The HEOM method has, in the last ten years, grown in popularity and applicability. The package we present in this work \cite{qutipweb1,qutipweb2,qutipweb3} captures most of this general applicability in being able to treat both bosonic and fermionic environments,  the modelling of multiple environments and time-dependant system Hamiltonians, as we have demonstrated above.  We have reproduced seminal results presented elsewhere, particularly the FMO treatment of [\onlinecite{Akihito09}] and the vibronic transport of [\onlinecite{schinabeck2}], and also demonstrated some novel applications, including dynamical decoupling with finite length pulses, demonstrating potential future applications in quantum control and noisy intermediate-scale quantum computing.

Future planned enhancements include full integration into QuTiP's quantum control and QIP libraries, support for time-dependent bath parameters (e.g., time-dependent chemical potential in fermionic systems), and support for solving problems with a system coupled to a fermionic and bosonic environment simultaneously within the HEOM description \cite{schinabeck}. For improved efficiency, further optimization of the fermionic solver construction (e.g., employment of the hermiticity relation discussed in [\onlinecite{hartle}] and [\onlinecite{HEOMQUICK}]) is planned, as well as  more robust and powerful parallel computing support, including GPU support.  

\acknowledgements
NL, TR and SC contributed equally to this work. The authors acknowledge discussions with Po-Chen Kuo, Mauro Cirio, Akihito Ishizaki, Xiao Zheng, Erik Gauger, Jakub Sowa, and Ken Funo.    
F.N. is supported in part by:
Nippon Telegraph and Telephone Corporation (NTT) Research,
the Japan Science and Technology Agency (JST) [via
the Quantum Leap Flagship Program (Q-LEAP), and
the Moonshot R\&D Grant Number JPMJMS2061],
the Japan Society for the Promotion of Science (JSPS)
[via the Grants-in-Aid for Scientific Research (KAKENHI) Grant No. JP20H00134],
the Army Research Office (ARO) (Grant No. W911NF-18-1-0358),
and 
the Asian Office of Aerospace Research and Development (AOARD) (via Grant No. FA2386-20-1-4069). F.N.~and N.L.~acknowledge support from the Foundational Questions Institute Fund (FQXi) via Grant No.~FQXi-IAF19-06. N.L. acknowledges support from JST
PRESTO through Grant No. JPMJPR18GC, and the Information Systems Division, RIKEN, for use of their facilities. PM performed part of this work as an International Research Fellow of the Japan Society for the Promotion of Science. DB acknowledges funding by the Australian Research Council (project numbers FT190100106, DP210101367, CE170100009). AP is grateful to MQCQE for supporting the opportunity to collaborate on this project as a visiting researcher.

\bibliography{references,rcrefs,more_refs}{}

\end{document}